\documentclass[fleqn,usenatbib]{mnras}

\usepackage{newtxtext,newtxmath}
\usepackage[T1]{fontenc}

\DeclareRobustCommand{\VAN}[3]{#2}
\let\VANthebibliography\thebibliography
\def\thebibliography{\DeclareRobustCommand{\VAN}[3]{##3}\VANthebibliography}

\usepackage{graphicx}	% Including figure files
\usepackage{amsmath}	% Advanced maths commands
\usepackage[toc,page]{appendix}
\usepackage{float}
\usepackage{longtable}
\usepackage{supertabular}
\usepackage{changepage}
\usepackage{subcaption}
\usepackage{natbib}
\usepackage{caption}
\usepackage{enumerate}
\usepackage{titlesec}
\usepackage{footnote}
\usepackage[section]{placeins}
\usepackage{hyperref}
\usepackage{siunitx}
\usepackage[utf8]{inputenc}
\usepackage[table]{xcolor}
\usepackage{xcolor}
\makeatletter
\patchcmd\@combinedblfloats{\box\@outputbox}{\unvbox\@outputbox}{}
\makeatother

\title[Spectral properties of NEOs with low-T$_J$]{Spectral properties of near-Earth objects with low-Jovian Tisserand invariant}
\author[N. G. Simion et al.]{N. G. Simion$^{1,2}$\thanks{Contact e-mail:\href{mailto:gabriel.nsimion@gmail.com}{gabriel.nsimion@gmail.com}}, 
	M. Popescu$^{1}$,
	J. Licandro$^{3,4}$,
	O. Vaduvescu$^{5,3}$,
	J. de Le\'on$^{3,4}$ and
	R. M. Gherase$^{1,6}$
	\\
	$^{1}$ Astronomical Institute of the Romanian Academy, 5 Cu\c{t}itul de Argint, 040557 Bucharest, Romania\\
	$^{2}$ Faculty of Physics, University of Bucharest, 405 Atomi\c{s}tilor str., M\u{a}gurele 077125, Ilfov, Romania \\
	$^{3}$ Instituto de Astrof\'{\i}sica de Canarias (IAC), C/V\'{\i}a L\'{a}ctea s/n, 38205 La Laguna, Tenerife, Spain\\
	$^{4}$ Departamento de Astrof\'{\i}sica, Universidad de La Laguna, 38206 La Laguna, Tenerife, Spain\\
	$^{5}$ Isaac Newton Group of Telescopes (ING), Apto. 321, E-38700 Santa Cruz de la Palma, Canary Islands, Spain\\
	$^{6}$ Faculty of Sciences, University of Craiova, Craiova, Romania
}

\date{Accepted XXX. Received YYY; in original form ZZZ}

\pubyear{2021}

\begin{document}
	\label{firstpage}
	\pagerange{\pageref{firstpage}--\pageref{lastpage}}
	\maketitle
	
	\begin{abstract}
		The near-Earth objects with low-Jovian Tisserand invariant ($T_J$) represent about 9\,per\,cent of the known objects orbiting in the near-Earth space, being subject of numerous planetary encounters and large temperature variations. We aim to make a spectral characterization for a large sample of NEOs with $T_J$\,$\leq$\,3.1. Consequently, we can estimate the fraction of bodies with a cometary origin. We report new spectral observations for 26 low-T$_J$ NEOs. The additional spectra, retrieved from different public databases, allowed us to perform the analysis over a catalogue of 150 objects. We classified them with respect to Bus-DeMeo taxonomic system. The results are discussed regarding their orbital parameters. The taxonomic distribution of low-$T_J$ NEOs differs from the entire NEOs population. Consequently, $T_J$$\sim$3 can act as a composition border too. We found that 56.2\,per\,cent of low-T$_J$ NEOs have comet-like spectra and they become abundant (79.7\,per\,cent) for T$_J$\,$\leq$\,2.8. 16 D-type objects have been identified in this population, distributed on orbits with an average T$_J$\,=\,2.65\,$\pm$\,0.6. Using two dynamical criteria, together with the comet-like spectral classification as an identification method and by applying an observational bias correction, we estimate that the fraction of NEOs with a cometary nature and H\,$\in$\,(14, 21)\,mag has the lower and upper bounds (1.5\,$\pm$\,0.15) and (10.4\,$\pm$\,2.2)\,per\,cent. Additionally, our observations show that all extreme cases of low-perihelion asteroids (q\,$\leq$\,0.3\,au) belong to S-complex.
	\end{abstract}
	
	
	
	%%%%%%%%%%%%%%%%%%%%%%%%%%%%%%%%%%%%%%%%%%%%%%%%%%
	
	%%%%%%%%%%%%%%%%% BODY OF PAPER %%%%%%%%%%%%%%%%%%
	
	\section{INTRODUCTION}
	
	At first glance, asteroids and comets are two different types of small objects, which can be found across the solar system. Their composition and orbital features are representative of the regions where they had been formed. Most comets are originating beyond the \emph{snowline} of early proto-Sun where water, carbon dioxide, or methane could condense. On the other hand, asteroids are inert bodies with more stable orbits on a large time scale and low content of volatile elements. Nevertheless, the discovery of active asteroids and dormant comets \citep[e.g.][and references there in]{2004come.book..659J,2015aste.book..221J} shows a continuum of properties between these two classes.
	
	A fraction of small bodies from the solar system has a perihelion distance q $\leq$ 1.3\,au. These are called near-Earth objects (NEOs). Currently, a number of 23\,873 NEOs is reported by the Minor Planet Center (accessed on October 31, 2020). The NEOs are a transient population since the average lifetime of an object is on the order of $\sim 10^7$\,yr \citep{2000Icar..146..176G}, which is way smaller than the solar system lifetime ($\sim 4.5 \times 10^9$\,yr). Thus, a source of replenishing is required, in order to explain their existence. Most of these replenished objects are asteroids originating from the main asteroid belt \citep[e.g.][]{1979Icar...37...96W, 2002Icar..156..399B, 2012Icar..217..355G}. They arrived in the near-Earth space due to a combination of the Yarkovsky effect and orbital resonances with Jupiter and Saturn. However, a fraction of NEOs displays dynamical features analogue to comets, while their visual aspect is similar to any other inert asteroid.
	
	On a large time scale ($\sim 10^4$\,yr) the periodic comets have unstable orbits, showing a chaotic evolution of their orbital elements \citep{1995A&A...299..288T}. As they approach the inner part of the solar system (a heliocentric distance smaller than 2\,au), they can display an outburst of gas and dust, known as \emph{cometary activity} (a coma followed by a tail). Consequently, their perihelion distance suffers frequent changes \citep{1992IAUS..152..269T}.  On a time scale of centuries, the close-encounters with planets may produce significant orbital instabilities for those bodies orbiting on comet-like orbits \citep[e.g.][]{2014Icar..234...66T}. Being composed of multiple types of small objects, with different origins, some NEOs could have come from outer regions as the Kuiper belt \citep{1997Icar..127...13L}. 
	
	A large fraction of comets does not show any activity, passing a period of dormancy. These dormant nuclei are objects covered (partially or entirely) by a refractory layer of dust and grains with a wide size distribution, formed after several perihelion passages  \citep[e.g.][]{1980ApJ...237..265B,1986ESASP.249..185R,1990A&A...237..524R,1996P&SS...44.1005B, 2002AJ....123.1039J}. These models explain the large differences in the apparent brightness of various comets in free sublimation state \citep{1989aste.conf..880W}. Once it reaches the surface, the gas flow exerts a pressure upon all small dust grains. If these are heavier than a threshold value, they will remain embedded on the surface, while all lighter grains are blown off. Eventually, after several perihelion passages, more and more heavy dust grains will be retained, giving birth to a mantle that ensure efficient insulation of deeper icy content. The scenario in which comets, after several perihelion passages, gradually consume their icy content and become extinct nuclei was also observed \citep{1985ASSL..115..149R}.
	
	The visual distinction between asteroids and active comets is almost trivial. However, in the case of small bodies with some comet-like orbital features and asteroid-like appearance, further investigations are necessary to reveal the possible cometary nature. From a dynamical viewpoint, these two types of objects have been distinguished for a long time based on their Jovian Tisserand invariant \citep[e.g.][]{1972BAICz..23....1K, 1992CeMDA..54..237K}, defined by Eq.~\ref{eq:Tisserand}:
	
	\begin{equation}
	\centering
	T_J = \frac{a_J}{a} + 2 \sqrt{(1-e^2) \frac{a}{a_J}} \cos(i) = \frac{2a_J}{q + Q} + \sqrt{\frac{8 q Q}{(q + Q) a_J}} \cos(i)
	\label{eq:Tisserand}
	\end{equation}
	where $a_J$ is the semi-major axis of Jupiter's orbit, while $e,~a,~i,~q,~Q$ are the eccentricity, semi-major axis, orbital inclination, perihelion and aphelion distances of the body's orbit. $T_{J}$ appears as a prime integral in the circular restricted three-body problem, involving the Sun, Jupiter, and a low-mass body.
	
	Comets with $T_J$\,<\,2 \& a\,<\,a$_{Nep}$ are labelled as Halley-type comets (nearly isotropic) \citep{1987A&A...187..899C} and those with $T_J$\,$\in$\,(2, 3) \& q\,<\,Q$_{Jup}$ are known as Jupiter Family Comets (JFCs - ecliptic comets), while the asteroids have $T_J$\,>\,3 \& q\,<\,Q$_{Jup}$ \citep{1972BAICz..23....1K}. JFCs are believed to come from regions beyond the orbit of Neptune, passing through the zone of the Jovian planets and finally being controlled only by Jupiter's gravity \citep{1980MNRAS.192..481F, 1997Icar..127...13L}. According to the numerical simulations of \cite{2002Icar..159..358F}, the JFCs spend $\sim$ a few 10$^3$\,yr in the near-Earth objects population, while their dynamical lifetime is on the order of  $\sim$ 3$\times$10$^5$\,yr. By comparing the observations with simulations, \citet{1994Icar..108...18L, 1997Icar..127...13L} had estimated that the physical lifetime of JFCs is somewhere between 3\,000 to 30\,000\,yr and concluded that the ratio of extinct to active ones is about 3.5.
	
	Although the value of $T_{J}$ can indicate whether or not the asteroid crosses Jupiter's orbit, this is not enough to produce orbital macroscopic changes, which are a characteristic of comets. \citet{2014Icar..234...66T} had developed a method to classify asteroids on cometary orbits (ACOs), based on orbital elements. The main advantage is that this method doesn't require any numerical time integration. Besides the Tisserand criterion, this algorithm rejects all objects in mean-resonant motion with Jupiter, those having large orbital uncertainties and large minimum orbital intersection distances (MOID) with the giant planets. We will refer to \emph{near-Earth asteroids on cometary orbits} (or NEACOs) for those following the \citet{2014Icar..234...66T} criterion and \emph{low-$T_J$ near-Earth objects} for the remaining ones, satisfying $T_{J}$\,$\leq$\,3.1.
	
	From physical viewpoint, active, dormant or extinct comet nuclei display C-, X- complex or B-, D-, T- type taxonomies, with featureless spectra \citep[e.g.][]{1989Sci...246..790V,2002AJ....123.1039J, 2003A&A...398L..45L, 2004come.book..223L, 2006AJ....132.1346C, 2012Icar..218..196D}. They are characterized by dark to reddish surfaces and low albedo (p$_v$\,<\,0.075), which represent signature of organic material on the surface \citep{1985Icar...64..503V,1987Icar...69...33H}.
	
	Because physical characterization is limited by observational opportunities, the number of NEOs with known physical properties is about 10\,per\,cent of those with known orbits \citep{2019A&A...627A.124P,2019Icar..324...41B}. The physical properties can be obtained through photometric, spectroscopic, polarimetric and radar observations. These are the most used methods to investigate the shape, taxonomic classification, albedo, surface composition and texture. 
	
	Multiple telescopic surveys have been performed over the years in order to find the fraction and properties of inactive comets. We will further highlight some of them. \citet{2005AJ....130..308F} had identified, based on mid-IR observations of 26 asteroids with T$_{J}$\,<\,3, a strong correlation between surface albedo and Tisserand parameter. Almost all studied asteroids with T$_{J}$\,<\,2.6 have had comet-like albedo (p$_{v}$\,<\,0.075) and, from those with T$_{J}$\,>\,3, only 10\,per\,cent were in this category. Based on this assumption, they estimated that about 4\,per\,cent of all known NEOs are extinct comets. 
	
	\citet{2008A&A...481..861L} analyzed the spectral properties of 41 asteroids with T$_{J}$\,<\,3. Within this sample, they noticed that objects in near-Earth orbits (q\,$\leq$\,1.3\,au) show a different taxonomic distribution, compared to other objects, in non-near-Earth orbits. They have reported that 35\,per\,cent of NEOs with T$_{J}$\,<\,3 show typical silicate absorption bands to 1 and 2\,$\mu m$, while only 4\,per\,cent of non-NEOs share similar features. They also found an anti-correlation between the Tisserand parameter and the spectral slope, with the reddest objects having the lowest T$_J$. 
	
	Another study that relates to physical properties of low-T$_J$ NEOs has been performed by \citet{2008Icar..194..436D}. Using a sample of 55 asteroids, they reported that (54\,$\pm$\,10)\,per\,cent of objects with T$_J$\,<\,3 have comet-like spectra or albedo and concluded that (8\,$\pm$\,5)\,per\,cent NEOs are inactive or dead comets.
	
	Recently, \citet{2018A&A...618A.170L} analyzed the spectra of 29 ACOs, which satisfy the dynamical criterion proposed by \citet{2014Icar..234...66T}. They found that all, except one spectrum having a low signal-to-noise ratio (SNR), show featureless spectra, similar to comets. They have also reported that all ACOs in long-period orbits (known as Damocloids) are D-types. This result is supported by the albedo values reported for a sample of 49 Jupiter Family Comets and 16  Damocloids by \citet{2016A&A...585A...9L}. They found that all ACOs (excepting three objects, for which the data had large error bars) have an albedo compatible with a cometary-like origin. However, when considering the general case of asteroids with T$_J$\,<\,3,  the recent study of \citet{2014ApJ...789..151K} report that 80\,per\,cent of them have the albedo $p_V$\,<\,0.1.
	
	In this paper, we seek to analyze the spectral properties of 150 near-Earth asteroids with T$_{J}$\,$\leq$\,3.1. Our study aims to find how many NEOs with low-T$_J$ values have spectral properties similar to those of cometary nuclei. Our results are analyzed in the context of other physical characteristics described in the literature. The article is organized as follows: the observational programme, the data reduction and the methods used to analyze the spectra are shown in Section~\ref{sec:methods}.  Then, Section~\ref{sec:results} presents the results for each  major spectral group. The findings are discussed in the context of other existing data in Section~\ref{sec:discussions}. Finally, Section~\ref{sec:conclusions} concludes the article.

	\section{OBSERVATIONS AND DATA ANALYSIS}
	\label{sec:methods}
	
	This section describes our observational programme, the complementary spectra retrieved from literature and the methods used to analyze the available spectral data. The catalogue of all 150 low-T$_J$ NEOs used in this study is shown in Table~\ref{neacos_table}. We also provide an introduction about the absolute magnitude distribution of the targeted population, the NEOs with T$_{J}$\,$\leq$\,3.1. 
	
	The boundary-value of T$_J$\,=\,3 is derived on the assumption of a hypothetical configuration of the solar system (consisting of only three orbiting bodies). However, because Jupiter's orbit is not circular ($e_{J}$\,=\,0.048) and the gravity of other planets is not negligible, we chose to use the threshold of T$_J$\,=\,3.1 to include the possible \emph{comets candidates} from the vicinity of T$_J$ \,=\,3, which may be affected by the perturbing effects of Jupiter.

	\subsection{The sample of NEACOs and low-T$_J$ NEOs}
	
	The opportunities to observe NEOs having T$_{J}$\,$\leq$\,3.1 and magnitudes brighter than $\sim$\,18.5\,mag (which is the upper limit for obtaining spectra using a 2-4\,m class telescope) are rare. Roughly, there are about five up to ten objects observable each semester. Mainly, this is caused by their small size and by their large eccentricity. Due to these eccentric orbits, most low-T$_J$ NEOs are orbiting at distances where they are not observable. Thus, in order to obtain a significant sample of spectral data, an observing programme must span many semesters.
	
	We started our observational project in August 2014, when the number of known NEOs was about 10\,000. Meanwhile, the new discoveries doubled this number and subsequently the one of NEOs with T$_{J}$\,$\leq$\,3.1, too. Also, other surveys which spectrally observed NEOs, had obtained data that are useful to our study. Hence, as a reference for this article we took the data available on October 31, 2020 from Minor Planet Center \footnote{\url{https://minorplanetcenter.net/}} website, namely the \emph{MPCORB.dat} file. It counts 23\,873 NEOs and it provides the following information for each object, the designation, the orbital elements, the epoch, the absolute magnitude, the mean daily motion  and some details related to orbit computation.
	
	Additionally, we used the spectral data available from SMASS-MIT-Hawaii Near-Earth Object Spectroscopic Survey (MITHNEOS MIT-Hawaii Near-Earth Object Spectroscopic Survey) programme \citep{2019Icar..324...41B} and from Modeling for Asteroids (M4AST) database \citep{2012A&A...544A.130P}. Most NEOs reported by these databases are easier to observe since they have large sizes (brightness) and are coming more frequent in Earth's vicinity. However, within our observational programme, we were focused on spectral studies of fainter NEOs, which are not detectable for most of their orbital period.

	We have generated the list of low-T$_J$ NEOs by calculating T$_J$ (using Eq.\,\ref{eq:Tisserand}) for all NEOs from the  \emph{MPCORB.dat} file and choosing those with T$_J$\,$\leq$\,3.1. As of October 2020, a total of 2\,179 asteroids were satisfying this condition, which represent $\sim$\,9\,per\,cent of the total number of NEOs. Over the years, we noted that this fraction didn't change with the discovery rate.
	
	The astronomers Gonzalo Tancredi and Silvia Martino have kindly provided us (personal communication) the updated list (as of October 2020) of asteroids on cometary orbits, which follow the criteria described by \citet{2014Icar..234...66T}. From this list, we have selected those objects orbiting in the NEOs population. Thus, we found 76 NEACOs, which represent about 0.32\,per\,cent of the total NEOs.
	
	\begin{figure}
		\centering
		\includegraphics[width=8.77cm, height = 7.2cm]{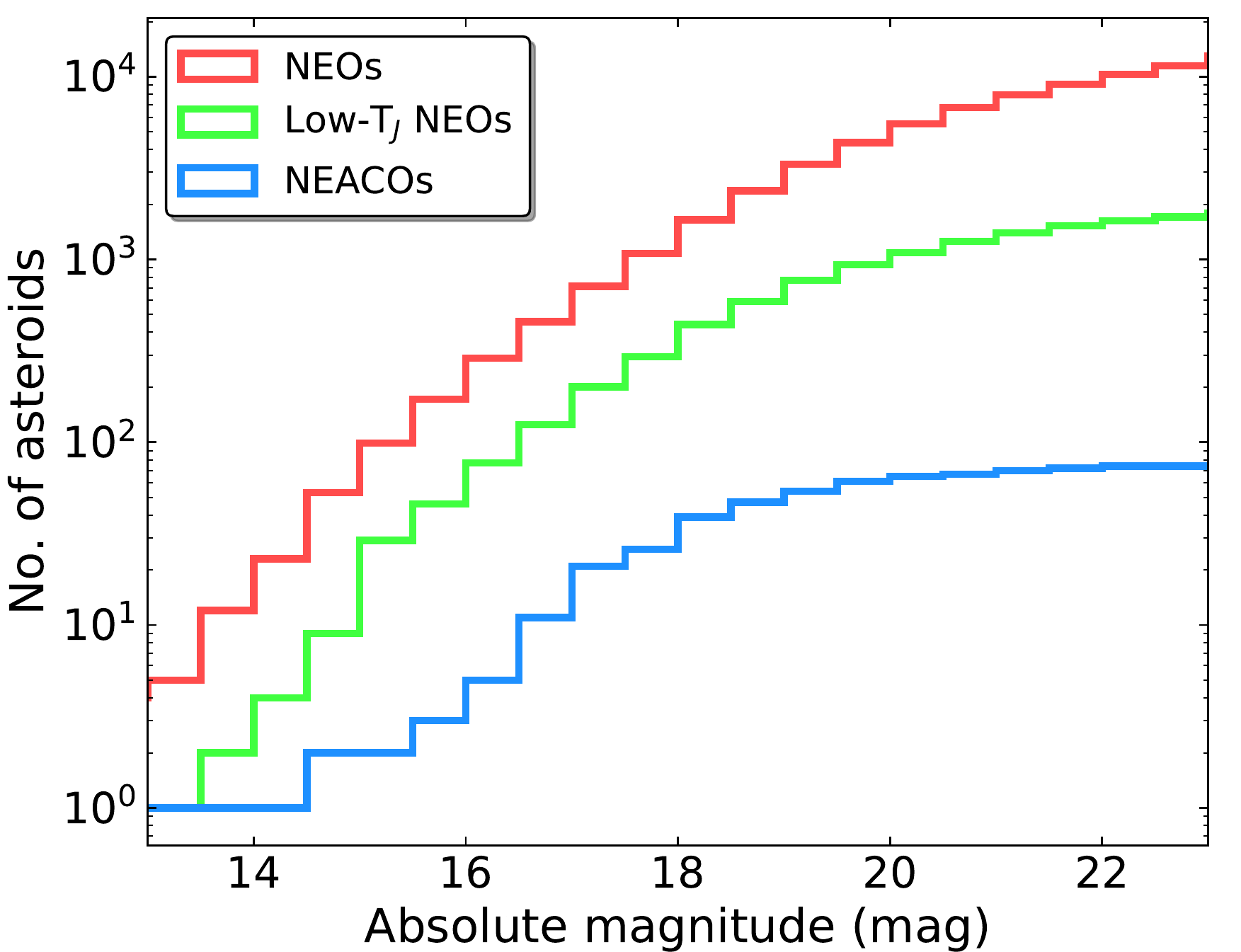}
		\caption{The absolute magnitude ($H$) cumulative distributions of all discovered NEOs (until October 31, 2020 - red), of all low-T$_J$ NEOs (green) and of all NEACOs (blue), following the definitions presented in the text . The horizontal axis is limited to (13, 23)\,mag interval to outline the relevant trend of the curves. A bin width of 0.5\,mag has been used for all three distributions.}
		\label{Fig:CumHNEA}
	\end{figure}
	
	The cumulative distributions with respect to absolute magnitude (Fig.~\ref{Fig:CumHNEA}) can indicate the discovery completeness \citep[e.g.][]{2002aste.book...71J, 2015Icar..257..302H} and are intricately linked to the size-frequency distribution of each of these populations. However, the curves shown in Fig.~\ref{Fig:CumHNEA} are modelled by observational biases \citep{2002aste.book...71J}, since the probability to observe (both for discovery and for spectral characterization) an asteroid depends strongly on the size, the albedo and its orbit.
	
	To verify whether or not the properties of NEOs, low-T$_J$ NEOs and NEACOs are statistically distinguishable, we have applied the Kolmogorov-Smirnov test (K-S test). This calculus is motivated because these subgroups are a small fraction of the NEOs population, and they have different orbits and various origins. The test takes two cumulative distribution functions (CDFs) as input and returns two probabilities, \emph{D-statistic} and \emph{p-value}. The first one is the maximum difference between the input cumulative functions. The second one is the probability that the maximum difference between other two samples, randomly chosen from the hypothetical common distribution of the input sets, is greater or equal to  \emph{D-statistic}.
	
	In order to apply the K-S test, we had first calculated the normalized cumulative distribution for each sample set. Only those objects having H\,$\in$\,(14, 21)\,mag were kept and grouped in a magnitude grid of 0.5\,mag step. The K-S test has been performed by using the \emph{SciPy Statistics} package \citep{2020SciPy-NMeth} from Python. In this module, an implemented function (\emph{ks\_2samp}) for this test can be found.
	
	For the population of NEOs (red line) and low-T$_J$ NEOs (green line), the K-S test returns a \emph{p-value} of 0.99 and a \emph{D-statistic} of 0.14, meaning that these two sample sets aren't distinguishable statistical-wise. Meanwhile, the KS-test between NEOs and NEACOs (blue line) gives a \emph{p-value} of 0.92 and a \emph{D-statistic} of 0.21. Since the \emph{p-value} is nearly unity in both cases, the K-S test revealed that the sample of low-T$_J$ NEOs and NEACOs are statistically linked with the entire NEOs population, describing the same absolute magnitude trend. In other words, the comet-like dynamical regions from the near-Earth space have no preferences for any particular values of absolute magnitude.

	\begin{table*}
		\caption{Summary of observational circumstances: asteroid designation, telescope/spectrograph used, diffraction instrument, date, the start time of each observation, exposure time, airmass, solar analogue and the references. The data marked as [1] or [2] in the \emph{Ref.} column were also reported by \citet{2019A&A...627A.124P} and \citet{2018A&A...618A.170L}.}
		\setlength\arrayrulewidth{0.5pt}
		\centering
		\rowcolors[]{1}{gray!0}{gray!7}
		\begin{tabular}{p{1.5cm} c c c p{1.2cm} c p{0.8cm} c c}
			\hline
			\rowcolor{gray!30}
			Object & Tel.-Instrum. & Grism/Prism, Slit & Date & UT$_{start}$ & t$_{exp}$(s) & Airmass & Solar Analogue & Ref.\\
			\hline 
			1580 & INT-RED+2 & R150V(0.4-0.92\,$\mu m$) & 26 May 2015 & \hfil 21:27 & 720 & \hfil 1.05 & HD109967 & [1]\\
			4401 & INT-RED+2 & R150V(0.4-0.92\,$\mu m$) & 13 Mar 2015 & \hfil 02:26 & 2700 & \hfil 1.1 & HD100044 & [1]\\
			9400 & INT-RED+2 & R150V(0.4-0.92\,$\mu m$)  & 31 Aug 2015 & \hfil 04:29  & 3$\times$300  & \hfil 1.095  & HD220764 & -\\
			112221 & NOT-ALFOSC & Grism\#4(0.5-0.92\,$\mu m$),1.8\,arcsec  & 04 Jun 2019 & \hfil 22:14 & 3$\times$450 & \hfil 1.45 &  SA107684  & -\\
			214088 & INT-RED+2 & R150V(0.4-0.92\,$\mu m$)  & 29 Dec 2014  & \hfil 01:19  & 3$\times$300  & \hfil 1.095  & HD31867 & -\\
			& INT-RED+2 & R150V(0.4-0.92\,$\mu m$)  & 06 Feb 2015  & \hfil 00:12  & 3$\times$1000  & \hfil 1.26  & HD283886 & -\\ 
			& IRTF-SpeX & low-res(0.8-2.5\,$\mu m$),0.8\,arcsec  & 07 Feb 2015  & \hfil 06:24 & 16$\times$120  & \hfil 1.021 & HD31867 & -\\
			248590 & IRTF-SpeX & low-res(0.8-2.5\,$\mu m$),0.8\,arcsec  & 09 Apr 2016 & \hfil 05:51 & 6$\times$120 & \hfil 1.431 & HD29714 & [2]\\
			276049 & INT-RED+2 & R150V(0.4-0.92\,$\mu m$) & 03 Sep 2014 & \hfil 00:15 & 900 & \hfil 1.05 & HD216516 & [1]\\
			285944 & INT-RED+2 & R150V(0.4-0.92\,$\mu m$) & 21 Aug 2014 & \hfil 01:50 & 840 & \hfil 1.18 & HD239928 & [1]\\
			293054 & INT-RED+2 & R150V(0.4-0.92\,$\mu m$)  & 21 Jul 2015  & \hfil 01:19  & 4$\times$1000  & \hfil 1.155  & HD131715 & -\\
			355256 & NOT-ALFOSC &Grism\#4(0.5-0.92\,$\mu m$),1.8\,arcsec  & 04 Jun 2019 & \hfil 21:40 & 3$\times$600 & \hfil 1.33 &  SA107684  & -\\
			413192 & INT-RED+2 & R150V(0.4-0.92\,$\mu m$)  & 29 Dec 2014  & \hfil 06:16  & 3$\times$660  & \hfil 1.23  & HD100044 & -\\
			414287 & INT-RED+2 & R150V(0.4-0.92\,$\mu m$)  & 02 Sep 2014  & \hfil 05:32  & 3$\times$900  & \hfil 1.013  & HD9761 & -\\
			416071 & INT-RED+2 & R150V(0.4-0.92\,$\mu m$)  & 06 Feb 2015  & \hfil 01:15  & 3$\times$800  & \hfil 1.42  & HD72911 & -\\
			& IRTF-SpeX & low-res(0.8-2.5\,$\mu m$),0.8\,arcsec  & 07 Feb 2015 & \hfil 09:27 & 16$\times$120  & \hfil 1.244 & HD72911 & -\\
			417264 & INT-RED+2 & R150V(0.4-0.92\,$\mu m$)  & 06 Feb 2015  & \hfil 23:58  & 3$\times$1000  & \hfil 1.20  & HYADE64 & -\\
			& IRTF-SpeX & low-res(0.8-2.5\,$\mu m$),0.8\,arcsec  & 07 Feb 2015 & \hfil 07:22 & 24$\times$120 & \hfil 1.062 & HD31867 & -\\
			429584 & INT-RED+2 & R150V(0.4-0.92\,$\mu m$) & 06 Feb 2015 & \hfil 06:08 & 2700 & \hfil 1.49 & HD110485 & [1]\\
			430439 & INT-RED+2 & R150V(0.4-0.92\,$\mu m$)  & 27 May 2015  & \hfil 01:51  & 4$\times$1200  & \hfil 1.05  & HD133623 & -\\
			433992 & IRTF-SpeX & low-res(0.8-2.5\,$\mu m$),0.8\,arcsec  & 14 May 2015 & \hfil 09:59 & 24$\times$120 & \hfil 1.073 & HD138159 & -\\
			& INT-RED+2 & R150V(0.4-0.92\,$\mu m$)  & 23 May 2015  & \hfil 23:20  & 2$\times$900,1$\times$327  & \hfil 1.497  & HD124446 & -\\
			442037 & INT-RED+2 & R150V(0.4-0.92\,$\mu m$)  & 21 Jul 2015  & \hfil 22:39  & 3$\times$200,2$\times$300  & \hfil 1.28  & SA93101 & -\\
			450160 & IRTF-SpeX & low-res(0.8-2.5\,$\mu m$),0.8\,arcsec  & 09 Apr 2016 & \hfil 07:11 &  4$\times$120 & \hfil 1.109 & HD89165 & -\\
			466130 & IRTF-SpeX & low-res(0.8-2.5\,$\mu m$),0.8\,arcsec  & 09 Mar 2016 & \hfil 14:54 & 13$\times$120  & \hfil 1.22 & HD142801 & -\\
			1998 GL10 & INT-RED+2 & R150V(0.4-0.92\,$\mu m$)  & 12 Mar 2015  & \hfil 02:31  & 3$\times$900 & \hfil 1.346  & HD91658 & -\\
			2008 JY30 & INT-RED+2 & R150V(0.4-0.92\,$\mu m$) & 27 Dec 2015 & \hfil 20:24 & 4800 & \hfil 1.15 & HD218633 & [1]\\
			2011 YB40 & INT-RED+2 & R150V(0.4-0.92\,$\mu m$)  & 25 Dec 2015  & \hfil 22:56  & 4$\times$1200 & \hfil 1.44  & HD89084 & -\\
			2015 CA1 & INT-RED+2 & R150V(0.4-0.92\,$\mu m$)  & 12 Mar 2015  & \hfil 03:20  & 3$\times$1000 & \hfil 1.32  & HD118928 & -\\
			2015 XB379 & INT-RED+2 & R150V(0.4-0.92\,$\mu m$)  & 25 Jan 2016  & \hfil 00:35  & 3$\times$900 & \hfil 1.062  & HD76617 & -\\
			& IRTF-SpeX & low-res(0.8-2.5\,$\mu m$),0.8\,arcsec  & 09 Mar 2016 & \hfil 07:03 & 10$\times$120  & \hfil 1.087 & HD76617 & -\\
			333P & IRTF-SpeX & low-res(0.8-2.5\,$\mu m$),0.8\,arcsec  & 09 Mar 2016 & \hfil 05:55 &  16$\times$120 & \hfil 1.26 & HD76617 & -\\
            & WHT & R400V(0.5-0.92\,$\mu m$),1.8\,arcsec & 18 Aug 2016  & \hfil 01:39 & 5$\times$600 & \hfil 1.16  & SA115271 & -\\
			\hline
		\end{tabular}
		\label{Obs. circumstances}
	\end{table*}
	
	The cumulative distribution of NEACOs is bent for bright and faint absolute magnitudes. The large object (3552) Don Quixote, with an absolute magnitude $H$ of 12.96\,mag does not follow the trend, being much larger than expected from these distributions (by extrapolating the NEACOs curve). For absolute magnitudes fainter than 21\,mag, the number of NEACOs drops because of the observational bias. Generally, these bodies are observable with ground-based telescopes for periods of days, when they come very close to Earth. We note that most low-T$_J$ NEOs with H\,>\,21\,mag ($\sim$\,93\,per\,cent) have uncertain orbits and were not classified as NEACOs.

	\subsection{New observations and their data reduction}
	
	We obtained new observations of NEOs with T$_{J}$\,$\leq$\,3.1 during August, 2014 - June, 2019, with telescopes from 2-4\,m aperture class. The visible spectra (0.4\,-\,0.9)\,$\mu m$ were obtained using the 2.5-m Issac Newton Telescope (INT), the 2.5-m Nordic Optical Telescope (NOT) and the 4.2-m William Herschel Telescope (WHT). All of them are located at the \emph{El Roque de los Muchachos Observatory} (ORM) in La Palma, Canary Islands (Spain), at an altitude of 2\,382\,m. For the NIR counterpart, we used the NASA 3.0-m Infrared Telescope Facility (IRTF - Mauna Kea, USA). Observational circumstances for all objects are presented in Table~\ref{Obs. circumstances}. 
	
	In the case of targets observed with INT we chose to use the IDS long-slit spectrograph, in a low-resolution mode. It has been equipped with the red-sensitive \emph{RED+2} CCD camera (4096 $\times$ 2048 pixels) as the detector. This provides a spatial scale of 0.44\,arcsec pixel$^{-1}$, with 15\,$\mu m$ pixel$^{-1}$. As diffraction instruments, IDS contains a set of 16 available gratings from which we chose R150V. With a dispersion of 271.3\,$\si{\angstrom}$\,mm$^{-1}$, it has a maximum efficiency of 65\,per\,cent. From our previous experience, this configuration has proven to be excellent for asteroids observation \citep{2019A&A...627A.124P}. A set of three to five images per target had been taken, with an individual integration time ranging from 300 to 1\,500 seconds (depending on the magnitude). 
	
	A visible spectrum of 2007 VA85 (recently reclassified as 333P/LINEAR) has been obtained using the ACAM (Auxiliary-port CAMera) instrument, which is mounted permanently at a folded-Cassegrain focus of the WHT. The 400-lines mm$^{-1}$ transmission VPH (Volume Phase Holographic) prism and a slit width of 1.8\,arcsec allows a resolution of $\sim$ 400. 
	
	Other two visible spectra were obtained using the ALFOSC (Alhambra Faint Object Spectrograph and Camera) instrument mounted on the NOT. The setup for this spectrograph includes the Grism\_\#4 together with a slit of 1.8\,arcsec width \citep[e.g.][]{2019CoSka..49..532K}.
	
	The NIR observations were performed with the SpeX instrument from IRTF \citep{2003PASP..115..362R}. This is a medium-resolution spectrograph, available since May 2000 and upgraded in August 2014. The CCD camera was also replaced in 2014, with the new Teledyne Hawaii-2RG detector (2048 $\times$ 2048 pixels). Each individual pixel has 18\,$\mu m$, equivalent to a spatial scale of 0.1\,arcsec pixel$^{-1}$. We have used this spectrograph in a low-resolution mode, covering the (0.8\,-\,2.5)\,$\mu m$ wavelength interval and a slit of 0.8 $\times$ 15\,arcsec. The observations have been performed in remote mode from \emph{Remote Observation Center in Planetary Sciences} (ROC) - Bucharest, Romania. The spectra of our targets (including the solar analogues) have been obtained alternatively on two separate locations on the slit, denoted A and B  (called an AB cycle), following the \emph{nodding procedure}.  The exposure time for an individual image of asteroids spectrum was 120\,s. Depending on the target magnitude, a number of  4-25 AB cycles were performed.
	
	The main observational constraints were that each NEO has its zenith angle z\,<\,60$^\circ$ (which corresponds to an airmass less than 2) and to avoid the galactic plane due to the high density of stars in the field. The minimization of errors introduced by differential refraction produced by the atmosphere was possible by orienting the slit along the parallactic angle. By orienting the slit in this way, it will be always parallel to the direction in which light is dispersed.
	
	The light reflected by an asteroid contains spectral signatures of both, the object and the Sun. As it passes through the Earth's atmosphere it includes also some specific features. The recorded light is modulated also by the transfer function of the instrument. To obtain the relative reflectance of our asteroids, we took spectra of several analogue standard stars for every observed target. The solar analogue is a star with the same spectral class as the Sun (G2V). In order to be used, they must be observed in the same conditions as the asteroid (the most important factor being the airmass). Thus, we divided the visible or NIR spectra of all samples by those of their corresponding solar analogue.
	
	In order to reduce the uncertainties introduced by the solar analogues, we observed before or after each asteroid observation a G2V star in the apparent vicinity of it. These have been selected from SIMBAD Astronomical Database - CDS (Strasbourg) \footnote{\url{http://simbad.u-strasbg.fr/simbad/}}, using the spectral type as the criterion for querying. In addition to these stars, which are mostly classified based on their colour indexes, during each observing night we have obtained spectra of at least one well known solar analogue (those that are commonly used by various studies which report asteroids spectra). For example, a list is shown in Table 1 of \citet{2020ApJS..247...73M}. In order to remove the outliers (miss-identification of solar analogues), we have compared between them all solar analogues observed during one night, and discarded those which show a profile different than most stars observed during the night. However, there is still an uncertainty of asteroid spectral slope on the order of $\sim$\,0.3\,per\,cent\,(1000\,$\si{\angstrom}$)$^{-1}$, due to solar analogue observation.
	
	The data reduction follows the same steps as the ones described by our team in previous articles, where the obtained data have been reported \citep{2014A&A...572A.106P,2019A&A...627A.124P, 2019CoSka..49..532K}. Firstly, we inspect all row spectra in order to avoid any artefacts. Then, the pre-processing of CCD images consists of bias and flat-field corrections (the flat fields were obtained using the internal lamps of each telescope/instrument, which are dedicated to this purpose). The extraction of 1D-spectrum from images has been made with the {\small IRAF - APALL} package for the  INT, NOT and WHT observations. The {\small GNU OCTAVE} software package \citep{octave} was used to create scripts for {\small IRAF} \citep{1986SPIE..627..733T}, in order to perform all these tasks. The wavelength map was determined using the emission lines from the lamps available for each set of observations.
	
	To reduce the IRTF spectra, we have used the {\small SPEXTOOL} pipeline \citep{2003PASP..115..362R, 2003PASP..115..389V, 2004PASP..116..362C}. Firstly, the bias and flat-field corrections are applied. To eliminate the sky contribution, the A and B images are subtracted, and the resulting A-B area is added to obtain a final image. The wavelength calibration is made using the emission lines of an argon lamp. The possible wavelength shifts (which generate the so-called \emph{heart beats} -- the reflectance spectral curve shows a positive spike followed or preceded by a negative one because of the wavelength miss-alignment) between the asteroid and the solar analogue are corrected using an IDL function included in the pipeline.
	
	\begin{table}
		\centering
		\setlength\arrayrulewidth{0.5pt}
		\caption{Statistic of the spectral data from each wavelength range studied in this work, for low-T$_J$ NEOs and NEACOs. A total of 150 spectra have been considered, 26 from our observational campaign and 124 retrieved from other public databases.}
		\rowcolors[]{1}{gray!0}{gray!7}
		\begin{tabular}{ p{1.67cm} | c c c}
			\hline
			\rowcolor{gray!30}
			Spectral range & VIS\,{\scriptsize (0.4-0.9)\,$\mu m$} & NIR\,{\scriptsize (0.9-2.5)\,$\mu m$} &\,VNIR {\scriptsize(0.4-2.5)\,$\mu m$}\\
			\hline
			Low-T$_J$ NEOs & 42 & 85 & 23 \\
			NEACOs & 3 & 3 & 1 \\
			\hline
		\end{tabular}
		\label{stats of spectra}
	\end{table}

	\subsection{Methods}
	
	The information regarding surface composition of an asteroid or comet can be obtained through spectroscopic and spectro-photometric observations, in the visible and NIR spectral range (abbreviated VNIR spectrum). A detailed analysis can be done when the full VNIR spectrum is available.
	
	\begin{figure*}
		\centering
		\includegraphics[width = 17.4cm]{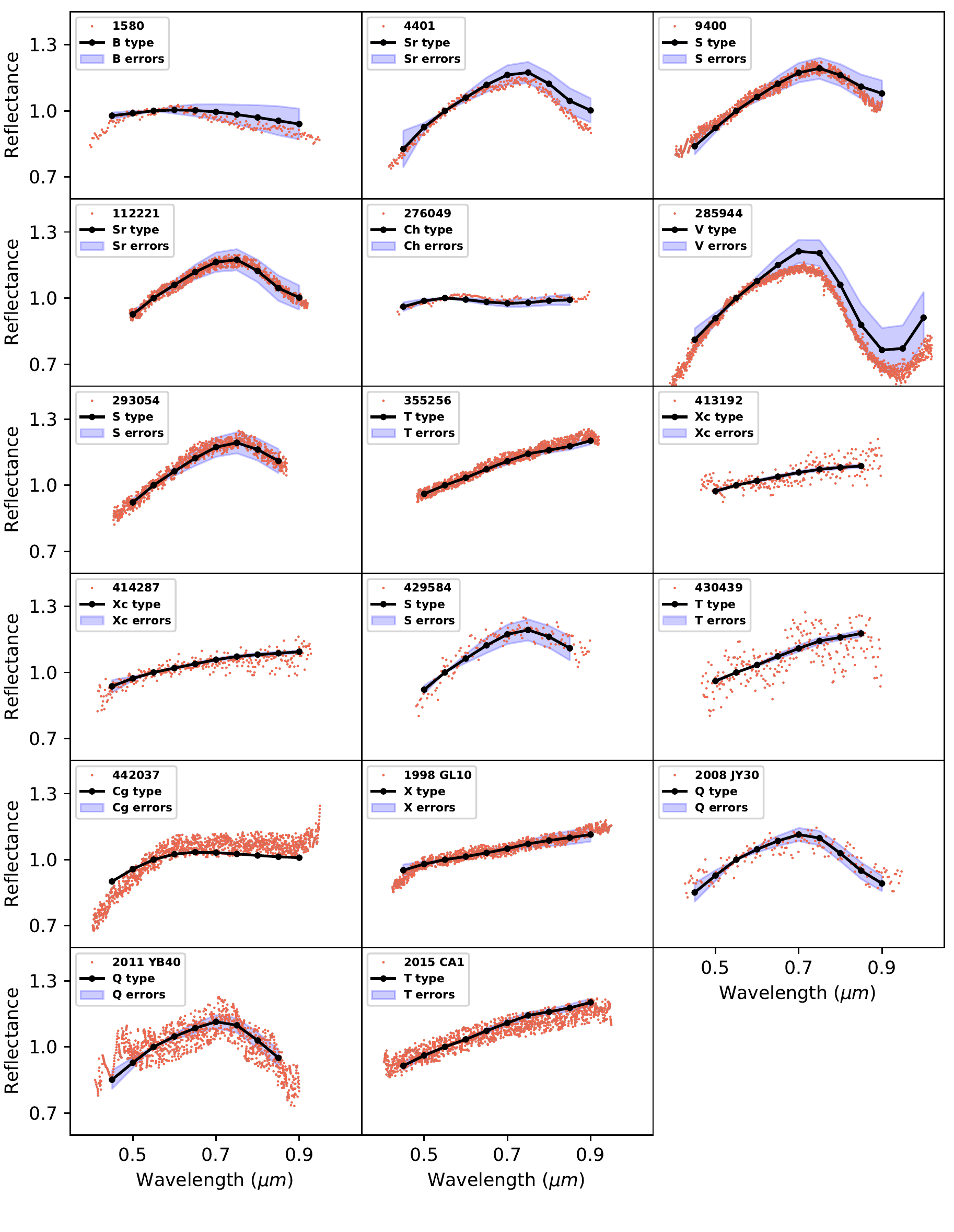}
		\caption{The visible spectra obtained with the INT and NOT telescopes in the framework of our observational programme. All of them have been normalized to 0.55\,$\mu$m. The black lines show the reference spectra, corresponding to each taxonomic type from \citet{2009Icar..202..160D} system and the blue shades are the standard deviations.}
		\label{Fig:vispectra}
	\end{figure*}
	
\begin{figure}
		\includegraphics[width = 8.65cm]{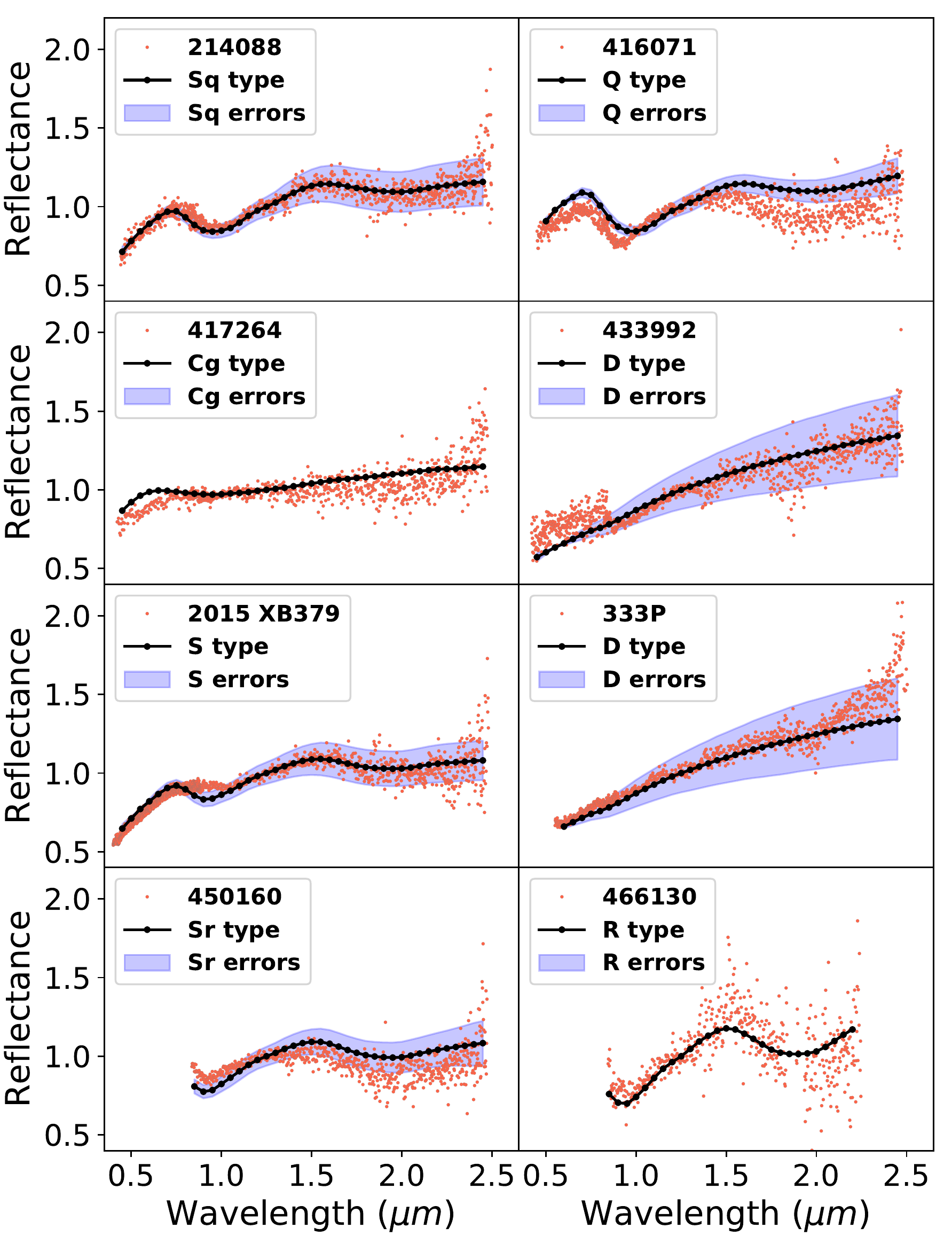}
		\caption{The VNIR (first six) and the NIR spectra (450160, 466130) obtained within our observational campaign. All of them have been normalized to 1.25\,$\mu m$. The black lines show the reference spectra, corresponding to each taxonomic type from \citet{2009Icar..202..160D} system and the blue shade is its standard deviation.}
		\label{Fig:vnirpectra}
	\end{figure}

	Within our observational programme, we performed spectroscopic observations for 26 low-T$_J$ NEOs. Most of them (17\,/\,26) have been measured in the visible range, while nine were observed only in NIR. For six objects (214088, 416071, 417264, 433992, 2015 XB379 and 333P) we recorded a complete VNIR spectrum (0.4\,-\,2.5)\,$\mu m$. 333P had been initially labelled as an asteroid (2007 VA85), but during its passage in early 2016, a cometary activity was observed \citep{2016MPEC....A..101L} and it was reclassified as an active comet. In Fig.\,\ref{Fig:vispectra} all visible spectra obtained are shown, while in Fig.~\ref{Fig:vnirpectra} are the NIR and VNIR observations.
	
	We have complemented the observed sample of spectral data used in this work with other existing observations. We searched the available spectral information from SMASS-MIT\footnote{\url{http://smass.mit.edu/}} and M4AST\footnote{\url{http://m4ast.imcce.fr/}} databases and found spectra for other 124 low-T$_J$ NEOs. Hence, the final catalogue (shown in Table~\ref{neacos_table}) contains data for 150 objects. The statistic is presented in Table\,\ref{stats of spectra}. 127 targets have observations only in one wavelength interval (42 visible and 85 NIR), while 23 NEOs have a full VNIR spectrum. The spectral results are discussed with respect to albedo and light-curve information. These were retrieved from MP3C\footnote{\url{https://mp3c.oca.eu/catalogue/index.htm}} and LCDB \footnote{\url{http://www.minorplanet.info/lightcurvedatabase.html}} databases, and the 
	NEODyS-2 -- Near Earth Objects Dynamic Site\footnote{\url{https://newton.dm.unipi.it/neodys/}}, which pointed to EARN database (no longer accessible) maintained by Dr. Gerhard Hahn for German Aerospace Center (DLR).

	\begin{table}
		\centering
		\setlength\arrayrulewidth{0.5pt}
		\caption{Summary of taxonomic classification for all objects observed within our observational programme. On the \emph{Taxonomy} column is the taxonomic type derived from our spectral observations.
			The \emph{Other tax.} column shows the previously known taxonomic classification according to the following references (1)-\citet{2019Icar..324...41B}, (2)-\citet{2006Icar..184..198S}, (3)-\citet{2002aste.book..669W}, (4)-\citet{2004Icar..170..259B}.}
		\rowcolors[]{1}{gray!0}{gray!7}
		\begin{tabular}{ p{1.5cm} p{0.7cm} p{1cm} p{1.6cm} p{1.4cm}}
			\hline
			\rowcolor{gray!30}
			Object & \hfil TJ & Taxonomy & Other tax. & Spec. range \\
			\hline
			1580 & \hfil 3.08 & \hfil B & \hfil B$^{(1)}$ & \hfil Vis. \\
			4401 & \hfil 3.06 & \hfil Q & \hfil Sr$^{(1)}$ & \hfil Vis. \\
			9400 & \hfil 2.95 & \hfil S & \hfil S$^{(4)}$ & \hfil Vis. \\
			112221 & \hfil 2.92 & \hfil Sr & \hfil - & \hfil Vis. \\
			214088 & \hfil 2.85 & \hfil Sq & \hfil Sq$^{(1)}$ & \hfil Vis.+Nir. \\
			248590 & \hfil 2.46 & \hfil X & \hfil - & \hfil Nir. \\
			276049 & \hfil 3.08 & \hfil Ch & \hfil C$^{(2)}$ & \hfil Vis. \\
			285944 & \hfil 3.06 & \hfil V & \hfil - & \hfil Vis.\\
			293054 & \hfil 2.94 & \hfil S & \hfil - & \hfil Vis. \\
			355256 & \hfil 2.77 & \hfil T & \hfil - & \hfil Vis. \\ 
			413192 & \hfil 2.78 & \hfil Xc & \hfil - & \hfil Vis. \\
			414287 & \hfil 2.62 & \hfil Xc & \hfil B$^{(1)}$ & \hfil Vis. \\
			416071 & \hfil 3 & \hfil Q & \hfil - & \hfil Vis.+Nir. \\
			417264 & \hfil 3 & \hfil Cg & \hfil - & \hfil Vis.+Nir. \\
			429584 & \hfil 3.07 & \hfil S & \hfil - & \hfil Vis. \\
			430439 & \hfil 2.93 & \hfil T & \hfil D$^{(3)}$ & \hfil Vis. \\
			433992 & \hfil 2.58 & \hfil D & \hfil - & \hfil Vis.+Nir. \\
			442037 & \hfil 2.82 & \hfil Cg & \hfil - & \hfil Vis. \\
			450160 & \hfil 2.7 & \hfil Sr & \hfil - & \hfil Nir. \\
			466130 & \hfil 2.39 & \hfil R & \hfil - & \hfil Nir. \\
			1998 GL10 & \hfil 2.79 & \hfil X & \hfil C|X$^{(1)}$ & \hfil Vis. \\
			2008 JY30 & \hfil 3.06 & \hfil Q & \hfil - & \hfil Vis. \\
			2011 YB40 & \hfil 2.99 & \hfil Q & \hfil - & \hfil Vis. \\
			2015 CA1 & \hfil 2.96 & \hfil T & \hfil - & \hfil Vis. \\
			2015 XB379 & \hfil 2.82 & \hfil S & \hfil - & \hfil Vis.+Nir. \\
			333P & \hfil 0.42 & \hfil D & \hfil - & \hfil Vis.+Nir. \\  
			\hline
		\end{tabular}
		\label{Tax_table}
	\end{table}

	The key step of our analysis was to assign a spectral type for each object, with respect to \citet{2009Icar..202..160D} taxonomic system (hereafter called Bus - DeMeo taxonomic system). This classification allows us to distribute objects with similar surface composition properties into common groups. Although a detailed mineralogical analysis requires high-resolution spectra, the taxonomic classification provides a first approach to constrain the composition features. Moreover, the classification can be performed even with low-SNR spectra. A taxonomic system consists of a set of standard  spectra, each of them labelled by a capital letter, known as type. Each label is assigned based on specific spectral features. In this way, such a system provides an alphabet for performing statistical studies. Those types which share similar features are grouped into complexes. {The Bus-DeMeo taxonomic system} is an update of the initial version of the Bus system (available only for visible part) \citep{1999PhDT........50B,2002Icar..158..146B}. It consists of 24 (25 according to the latest update) spectral types, defined over (0.45\,-\,2.45)\,$\mu m$ wavelength interval. Besides the spectral interval over which they were defined, there are some minor differences between the classes of Bus - DeMeo taxonomy and those of Bus taxonomy (these are highlighted by \citet{2009Icar..202..160D}. However, in order to perform our study, we need to work with an unique system, thus we used as references the classes defined in Bus\,-\,DeMeo taxonomic system.
	
	All spectra (visible, NIR and VNIR) were classified using the Modeling for Asteroids (M4AST) interface \citep{2012A&A...544A.130P}. The distances between the spectrum and the reference spectral types are computed based on the mean-squared differences. The first three matched classes, which correspond to the minimum distances, are returned. These were visually inspected and the class which best fitted the spectral features was selected (in most cases, this was the closest one in terms of mean-squared differences). When the NIR or the full VNIR spectrum (0.45\,-\,2.45)\,$\mu m$ was available, we also used for comparison the classification tool provided by SMASS-MIT. This tool computes the principal components and assigns a class based on the space defined by \citet{2009Icar..202..160D}, using 371 reference spectra. In most cases, the two methods gave the same classification. However, when the results provided by these tools didn't match or were ambiguous, we assigned the class which matches best the spectral features, by visual inspection. Table\,\ref{Tax_table} shows a summary of the taxonomic classification for all 26 low-T$_J$ NEOs observed in our programme.
	
	When the full VNIR spectrum is available, the spectral classification is more accurate. The merging of visible observations with the NIR counterparts was performed using the \emph{Concatenation} routine provided by M4AST. This implies the minimization of differences in the common wavelength interval, where in most cases was around 0.82\,-\,0.95\,$\mu m$. Some mismatches exist in the common interval. Most of them were outliers points, as a result of non-identical observational circumstances including airmasses, phase angles, solar analogues or various instruments \citep{2014A&A...572A.106P}. These are because different telescopes (or instruments) were required to obtain the visible and NIR part. For some objects, we've had several visible and NIR spectral curves. In order to obtain the best matches between them, every combination has been verified. We selected the spectra with the best matching in the common interval and with the highest signal-to-noise ratio (SNR).
	
	All spectra have been normalized. The normalization of visible spectral curves has been made by dividing them to the median reflectance between (0.5\,-\,0.6)\,$\mu m$.  For the normalization of NIR and VNIR data, the same procedure has been applied for the points between (1.2\,-\,1.3)\,$\mu m$. These two normalization procedures are equivalent with the normalization at 0.55\,$\mu m$, or respectively 1.25\,$\mu m$ but they are not sensitive to noise.
	
	The equivalent diameter (D$_{eq}$) of each object has been estimated using the formula (Eq.~\ref{bottkediam}) provided by \cite{1989aste.conf..524B}
	\begin{equation}
	D_{eq}(km) = \frac{1329}{\sqrt{p_v}} \cdot 10^{-0.2 \cdot H}
	\label{bottkediam}
	\end{equation}
	The absolute magnitudes (\emph{H}) were provided by MPCORB.dat file, while the albedo ($p_v$) was retrieved from the MP3C database{\bf \footnote{\url{https://mp3c.oca.eu/}}} and it is determined by WISE \citep{2011ApJ...741...68M, 2011ApJ...743..156M, 2014ApJ...792...30M, 2019PDSS..251.....M} and AKARI observations \citep[e.g.][and references there in]{2018A&A...612A..85A}. When $p_V$ was not available, we considered the average value for the corresponding taxonomic class \citep{2011ApJ...741...90M}. A comprehensive study of the correlation between albedos and taxonomic classification of NEOs was reported by \citet{2011AJ....142...85T}. They show that the S-complex, Q- and V-type NEOs have higher average albedos, compared to the main asteroid belt. This suggests that our size estimation may have an uncertainty up to $\sim$\,10\,per\,cent. However, the value they found for the C-complex asteroids is computed using only five albedo values, which are spread over a wide range. Thus, for consistency, we prefer to use only the average albedo values of \citet{2011ApJ...741...90M}.
	
	Another property that has been considered is the surface temperature. Since the NEOs on comet-like orbits have large eccentricities, they are the subject of large temperature variations between perihelion and aphelion passages. Hence, using the Eq.~\ref{Eq:temp} \citep{2009M&PS...44.1331B}, we have calculated the surface temperature reached at perihelion and aphelion. 
	
	\begin{equation}
	T (r) = \left[\frac{(1 - A)\cdot L_\odot}{16  \eta  \epsilon  \sigma  \pi  r^2} \right] ^ {\frac{1}{4}}
	\label{Eq:temp}
	\end{equation}
	where A is albedo, $L_\odot$ is solar luminosity (3.82\,$\cdot\,10^{26}$ W), $\eta$ is beaming factor (chosen as unity) \citep{1998AJ....115.1671C}, $\epsilon$ is asteroid's infrared emissivity (0.9), $\sigma$ is the Stefan-Boltzmann constant and r is the heliocentric distance (in meters).
	
	Last but not least, for cometary candidates spectra (belonging to C-, X- complexes, D-, T- types) we calculated the visible and NIR spectral slope following the definition shown by \citet{1996AJ....111..499L, 2002AJ....123.1039J}, over a wavelength interval identical with the one used by \citet{2018A&A...618A.170L}. This is a measure of reflectivity's variation over 1000\,$\si\angstrom$.
	
	\begin{equation}
	S' = \frac{1}{S(\lambda_{{\tiny norm}})} \cdot \frac{\Delta S}{\Delta \lambda}
	\end{equation}

	Because most comet-like spectra are featureless, we have approximated each spectrum with a first-order polynomial fit. In order to avoid the spectrum's strange artefacts or the possible thermal emission (in the NIR part), which could strongly influence the slope value, we have considered the (0.55, 0.85)\,$\mu m$ wavelength interval for the visible slopes and the (1, 1.75)\,$\mu m$ spectral range for the NIR slopes. These computations are performed by normalizing the spectrum at 0.55 and 1.2\,$\mu m$, respectively. In the above relation, S($\lambda_{norm}$) represents the reflectance to which the spectrum has been normalized, $\Delta S$ is the reflectance difference between $\lambda_{max}$ and $\lambda_{min}$ and $\Delta \lambda$ is the wavelength interval. We have been able to compute the visible slope for 32 objects, whereas the NIR ones were calculated for 51 objects.

	\section{RESULTS}
	\label{sec:results}
	
	Our approach to distinct spectra which could correspond to comet-like objects was to group all 25 taxonomic classes, defined within the Bus-DeMeo system, into four composition categories. A similar grouping has been done by \citet{2016AJ....151...11P}, who studied the physical properties of potentially hazardous asteroids (PHAs).
	
	Hence, the comet-like spectral group includes the C-complex (B, C, Cb, Cg, Ch, Cgh), similar to carbonaceous chondrite meteorites. It also includes the moderate to highly-slope spectra of T and D-type, which are considered among the most primitive bodies in the solar system and are supposed to contain organics and volatiles \citep[e.g.][]{2001Sci...293.2234H,2018MNRAS.476.4481B,2021Icar..36114349G}. In this group, we included the X-complex (X, Xe, Xk, Xc, Xn) which may represent various compositions and for which the albedo is needed to identify a possible cometary-like body. The silicate-like spectral group mainly consists of objects from the Q\,/\,S-complex (S, Sa, Sr, Sq, Sv) and some end-members like O, R and A-types. These have the 1 and 2\,$\mu m$ absorption bands as the main features, excepting A-type, which describes olivine-dominated asteroids and shows only the 1\,$\mu m$ feature. Another group consists of basaltic-like asteroids, corresponding to V-type spectra. Basaltic objects are one of the silicate end-member types (similar to O, R or A-types), but we decided to discuss them separately due to their specific nature. The last group has relatively rare spectra (denoted as \emph{miscellaneous}) of K and L-type. 
	
	\subsection*{Taxonomic distribution}
	
	\begin{figure}
		\centering
		\includegraphics[width = 8.8cm]{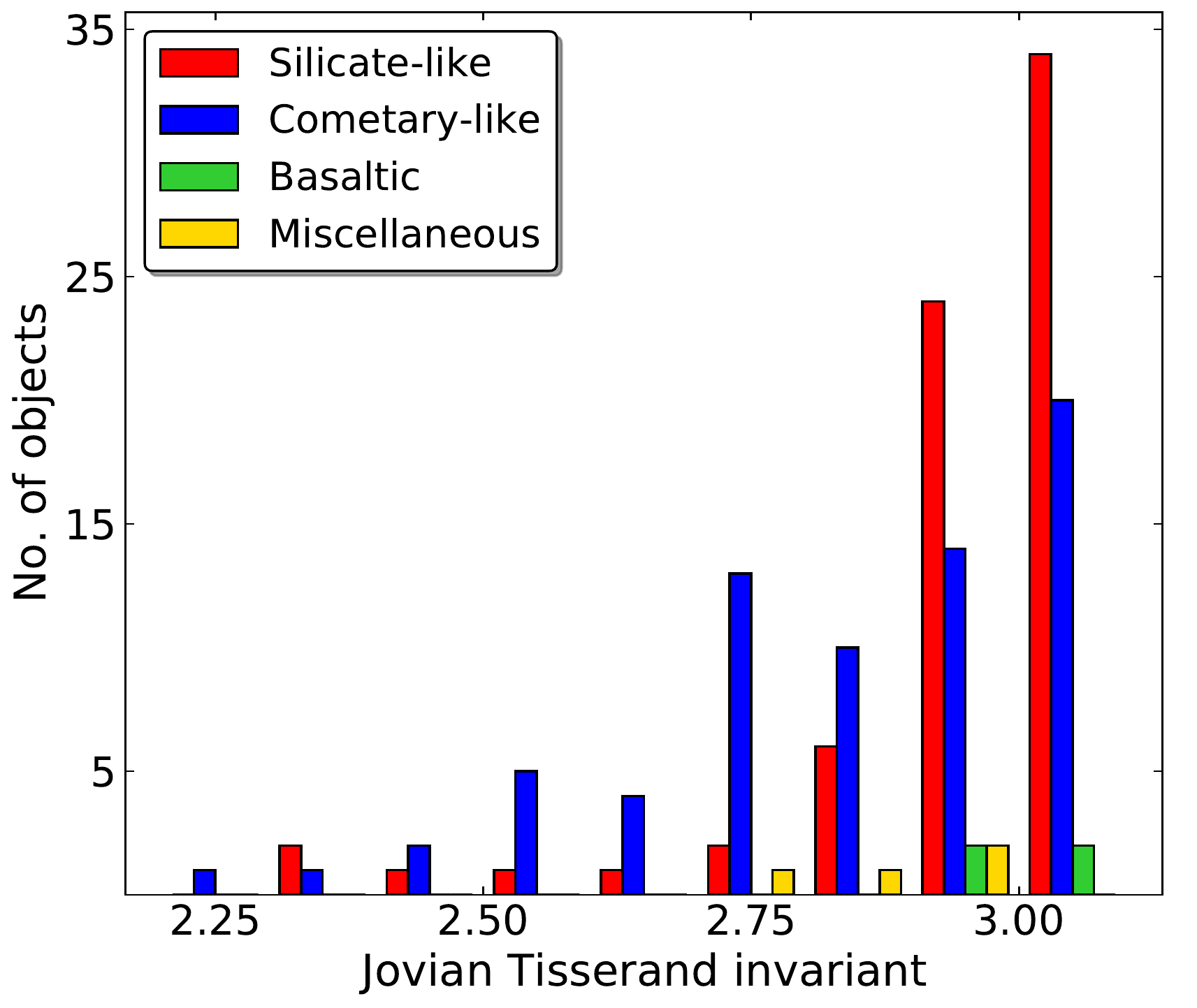}
		\caption{The taxonomic distribution illustrated by our entire catalogue (150 objects), relative to the Jovian Tisserand parameter (T$_{J}$). A bin width of 0.1 has been used. All objects have been grouped into four composition groups, based on their spectral features: silicate-like spectra (Q\,/\,S-complex, R-, A-type), cometary-like spectra (B\,/\,C-, X- complexes, D-, T- type), basaltic spectra (V-type) and miscellaneous types (K- and L-type).}
		\label{Fig:tax_all}
	\end{figure}
	
	The taxonomic distribution allows us to find the general picture of NEOs population in terms of composition \citep{1982Sci...216.1405G}.  Because the main asteroid belt is the principal source responsible for resupplying the NEOs population, the latter one displays the same diversity taxonomic-wise. Consequently, the NEOs population is dominated by Q\,/\,S-complex (including all sub-classes), representing more than half (5-10\,per\,cent Q-type and $\sim$\,50\,per\,cent S-complex). It is followed by B\,/\,C-complex ($\sim$\,\,15\,per\,cent) and X-complex (10\,per\,cent). Basaltic asteroids,  having V-type taxonomy, represent a fraction of $\sim$\,5\,per\,cent, while D-type objects make up to 3\,per\,cent \citep{2015aste.book..243B, 2019Icar..324...41B, 2019A&A...627A.124P}. These percentages are slightly different for NEOs with sizes below 300\,m \citep[e.g.][]{2018P&SS..157...82P, 2018MNRAS.477.2786P, 2018MNRAS.476.4481B, 2019AJ....158..196D}. In particular, \citet{2018MNRAS.476.4481B} noticed that D-types become more abundant in the population of smaller NEOs.
	
	The taxonomic distribution of our sample of 150 NEOs, compiled with respect to Jovian Tisserand invariant, is shown in Fig.\,\ref{Fig:tax_all} and the statistic of each composition group is presented in Table\,\ref{Table:taxonomy}. A relation between T$_{J}$ (a dynamical property) and the taxonomic type (physical feature) can be observed. This is highlighted by the ratio between the silicates and the comet-like bodies, which varies with T$_{J}$. The inversion of the composition ratio takes place around 2.8. Over 90\,per\,cent of Q\,/\,S-complex bodies are confined to T$_{J}$\,$\geq$\,2.8 and only 38\,per\,cent of low-T$_J$ NEOs with comet-like spectral type are characterized by T$_{J}$\,$\leq$\,2.8. 
	
	One can observe that in the T$_J$\,$\leq$\,3.1 region, the composition trend differs significantly from the corresponding distribution of the entire NEOs population. Within our data set, the comet-like group contains nearly half of all objects (71\,/\,150, representing 47.3\,per\,cent), while in the entire NEOs population they represent only $\sim$\,25\,per\,cent \citep{2019Icar..324...41B}. Hence, the T$_{J}$\,$\sim$\,3 threshold is also a border in terms of composition, and not only a dynamical one. We note that this distribution is based only on the observed data and it is influenced by the observational bias  (preference for observing the brightest objects, and those which are on favourable geometries of observations for longer time intervals). This effect is reviewed in Section 4.1.   
	
	\begin{table}
		\centering
		\setlength\arrayrulewidth{0.5pt}
		\caption{The number of objects with known spectral type from each composition group considered in this work, for low-T$_J$ NEOs and NEACOs.}
		\rowcolors[]{1}{gray!0}{gray!7}
		\begin{tabular}{l | c c c c}
			\hline
			\rowcolor{gray!30}
			Taxonomic group & Cometary & Silicate & Basaltic & Miscellaneous \\
			\hline
			Low-T$_J$ NEOs & 71 & 71 & 4  & 4 \\
			NEACOs & 7 & - & -  & - \\
			\hline
		\end{tabular}
		\label{Table:taxonomy}
	\end{table}

	\subsubsection*{The comet-like spectral group}
	
\begin{figure}
		\centering
		\includegraphics[width = 8.8cm]{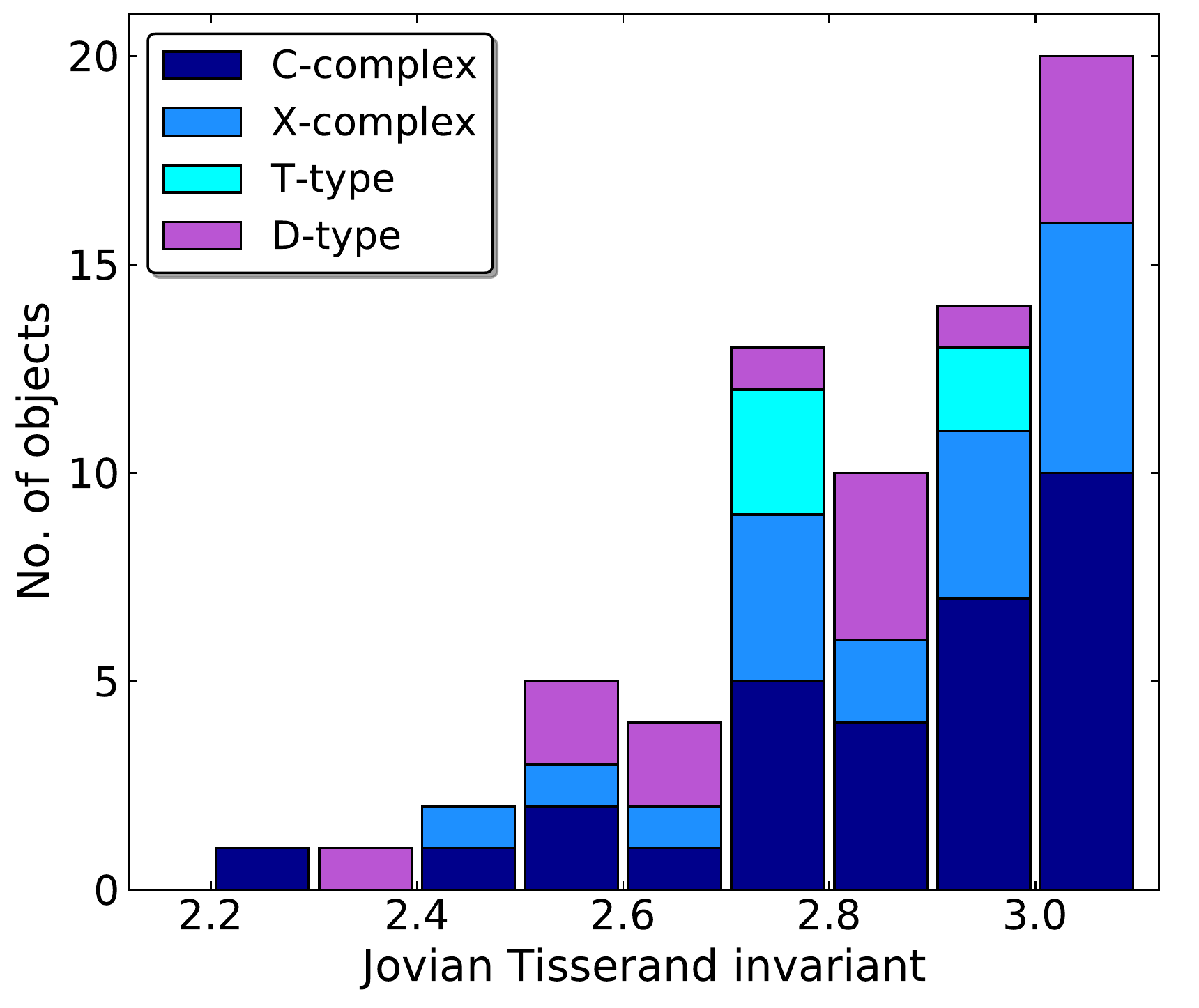}
		\caption{The taxonomic distribution of all 71 low-T$_J$ NEOs within the comet-like composition group, relative to their Jovian Tisserand invariant. A bin width of 0.5 has been used. The B\,/\,C-complex objects are in majority (31\,/\,71), followed by X-complex (19\,/\,71), D-type (16\,/\,71) and T-type (5\,/\,71).}
		\label{Fig:tax_tj_carb}
	\end{figure}

	The distribution of the spectral types within this group, relative to T$_J$ is shown in Fig.~\ref{Fig:tax_tj_carb}. It can be observed that the comet-like group is dominated by the B\,/\,C-complex (43.7\,per\,cent including 6 B-types, 16 C, 2 Ch, 2 Cb and 5 Cg). It is followed by the X-complex ($\sim$\,27\,per\,cent) having 12 X, 1 Xk and 6 Xc. Finally, the objects with highly-slope spectra make up to 29.6\,per\,cent, 16 D- and 5 T-type. A number of 40 (from a total of 71) NEOs from the comet-like spectral group have been observed only in the NIR range, while 19\,/\,71 have spectra only in the visible range. A full VNIR (0.4\,-\,2.5)\,$\mu m$ spectral characterization was retrieved for 12 of these objects. 
	
	Most spectral features of comet-like asteroids can be highlighted only by measurements in the visible range. For instance, Cg-type spectra present a drop of reflectivity below 0.55\,$\mu m$ \citep{2009Icar..202..160D}, while  Cgh\,/\,Ch-types are characterized by a broad and shallow absorption band (0.55\,-\,0.85\,$\mu m$, centred to 0.7\,$\mu m$). This feature marks the existence of hydrated minerals on the surface (i.e. minerals who suffered aqueous alteration) \citep{1989Sci...246..790V}. C-type spectra can be identified by flat or small slope profiles, similar with CI, CC or CM hydrated carbonaceous chondrite \citep{1985Icar...64..503V,2011Icar..212..180C, 2011Icar..216..309C}. 
	
	The B-types are the only ones characterized by a negative blue slope, which makes them easy to distinguish. The negative slope's trend continues in the NIR part. Finally, the end-members spectra of D- and T-types show featureless and highly slope profiles, in both visible and NIR regions. This is an indicator of organic composition and volatiles on the asteroid's surface \citep[e.g.][]{1980Natur.283..840G}, which is directly linked to regions where they had been formed. D-types are considered to have the most primitive surface composition, making them proper comet candidates in our analysis.
	
	The asteroids classified as X-complex objects show a broad distribution of surface composition and albedo, with spectra without absorption features and low to high slopes. Similar signatures were found to carbonaceous chondrites \citep{2005A&A...430..313B}, enstatite chondrites \citep{2009Icar..202..477V}, stony-iron and iron meteorites \citep{2010Icar..210..674O, 2011Icar..214..131F}. Yet, not all X-complex asteroids can have a comet-like nature. Thus, knowing of surface albedo becomes essential to reduce this uncertainty. Out of 19 X-complex objects, we retrieved the albedo value only for eight of them (2 X, 5 Xc, 1 Xk).
	
	We note in this group the case of (414287) 2008 OB9, for which the visible spectra and the NIR data does not match. We report an Xc-type classification, while the NIR observations (retrieved from SMASS-MIT database) show a blue slope, which corresponds to a B-type object. In this case, we have considered the classification derived based on our data. Another interesting case is of (276049) 2002 CE26. For this asteroid, we assigned a Ch-type classification, derived from spectral data in the visible range. Based on NIR and radar observation, \citet{2006Icar..184..198S} deduced that 276049 is a binary near-Earth object, with a primary of (3.5\,$\pm$\,0.4)\,km and a secondary of $\sim$0.3\,km. As for taxonomy, they have classified it as a C-type. Yet, the visible spectrum we have obtained (see Fig\,\ref{Fig:vispectra}) has the 0.7\,$\mu m$ feature, indicating the presence of hydrated minerals on the asteroid's surface.
	
	\begin{figure*}
		\centering 
		\begin{subfigure}{0.49\textwidth}
			\includegraphics[width = 8.87cm]{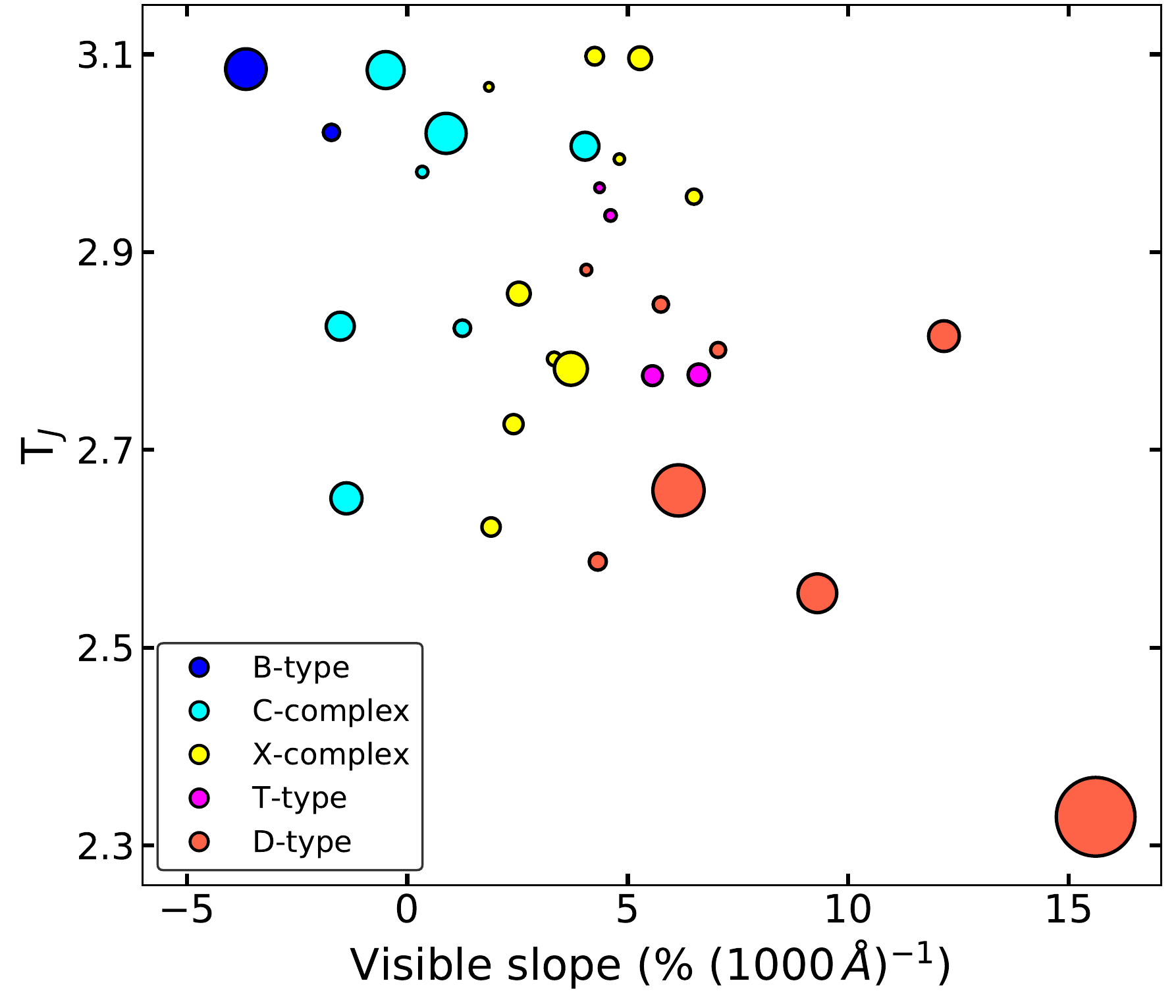}
			\caption{}
			\label{Fig:Vis_slopes}
		\end{subfigure}
		\begin{subfigure}{0.48\textwidth}
			\includegraphics[width = 8.87cm]{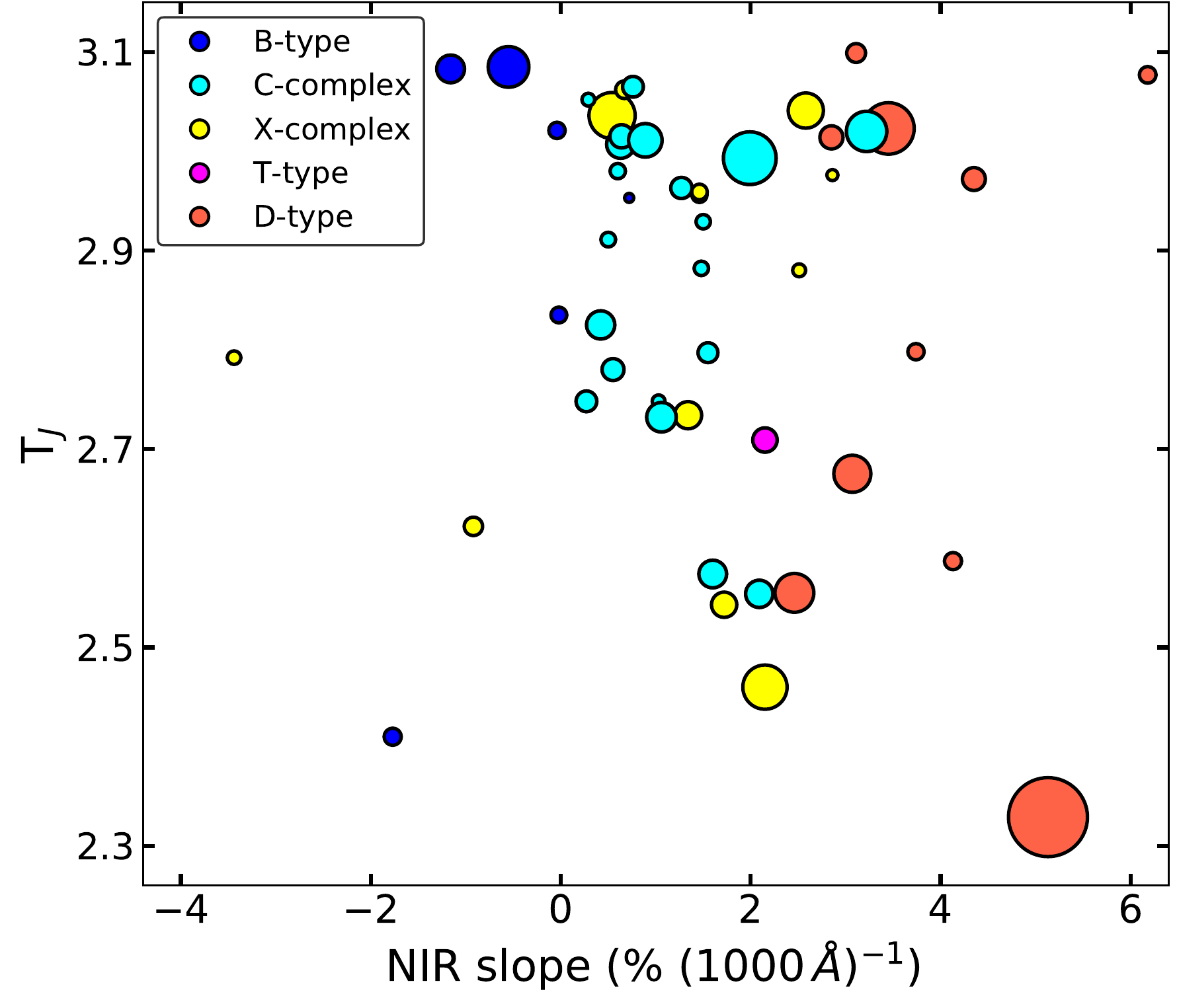}
			\caption{}
			\label{Fig:NIR_slopes}
		\end{subfigure}
		\caption{The distribution of visible (left panel) and NIR (right panel) slopes, relative to the Jovian Tisserand invariant for all low-T$_J$ NEOs with taxonomy from the cometary-like spectral group. The area of each data point is plotted as a function of the equivalent diameter (D$_{eq}$) of each corresponding object. Thus, the size difference between every two data points is proportional to the difference between objects' diameters. The size varies from 0.2 to 15.8\,km on the left panel and from 0.3 to 15.8\,km on the right panel. The largest value (15.8\,km) corresponds to (3552) Don Quixote (lower-right part of both figures).} 
	\end{figure*}
	
	Dynamically, almost all C-complex objects (30\,/\,31) within our set have large perihelia (q\,$\geq$\,0.7\,au) and eccentricities e\,$\in$\,(0.45, 0.76). The single extreme case is (170502) 2003 WM7 (e\,$=$\,0.88), who approach to the Sun at a minimum distance of 0.297\,au. Also, this asteroid is the subject of large temperature variations. Based on the retrieved albedo (p$_v$\,=\,0.05) and using Eq.\,\ref{Eq:temp}, we estimate a variation of $\Delta$T\,$\sim$\,400\,K between perihelion and aphelion passage. Similarly, D and T-type objects are divided as follow: 19\,/\,21 have q\,$\geq$\,0.75\,au and e\,$\in$\,(0.55, 0.7), while below q\,=\,0.5\,au there are two D-types with large eccentricities, namely 139359 (e\,$=$\,0.86, q\,$=$\,0.34\,au) and 401857 (e\,$=$\,0.855, q\,$=$\,0.4\,au).
	
	\begin{table*}
		\centering
		\setlength\arrayrulewidth{0.5pt}
		\caption{Summary of the average, median and standard deviation of T$_J$, visible and NIR slopes, for various categories of objects. The visible slopes were computed over the (0.55\,-\,0.85)\,$\mu m$ interval, normalized to 0.55\,$\mu m$, whereas the NIR ones were calculated for the (1\,-\,1.75)\,$\mu m$ interval and normalized to 1.2\,$\mu m$.  On the N$_{vis}$ and N$_{NIR}$ columns is the total number of objects within each category, considered for each spectral region (visible and NIR). Also, on the T$_J$ columns, the left values correspond to objects with visible slopes, while the right values are for objects with NIR slopes.}
		\rowcolors[]{1}{gray!0}{gray!7}
		\begin{tabular}{l c c c c c c c c c c c}
			\hline
			\rowcolor{gray!30}
			Type of objects & N$_{vis}$ & N$_{NIR}$ & S$_{vis}^{avg}$ &  S$_{vis}^{med}$ & S$_{vis}^{std}$ & S$_{NIR}^{avg}$ & S$_{NIR}^{med}$ & S$_{NIR}^{std}$ & T$_{J (vis|NIR)}^{avg}$ & T$_{J (vis|NIR)}^{med}$ & T$_{J (vis|NIR)}^{std}$ \\
			\hline
			Low-T$_J$ NEOs & 32 & 51 & 3.98 & 4.15 & 3.9 & 1.54 & 1.46 & 1.72 & 2.77\,|\,2.79 & 2.83\,|\,2.91 & 0.46\,|\,0.4 \\
			B\,/\,C-complex NEOs & 9 & 27 & -0.25 & -0.5 & 2 & 0.77 & 0.72 & 0.98 & 2.94\,|\,2.85 & 3\,|\,2.92 & 0.15\,|\,0.2 \\
			D-type NEOs & 9 & 11 & 8 & 7 & 3.6 & 3.88 & 3.74 & 1 & 2.43\,|\,2.59 & 2.65\,|\,2.8 & 0.73\,|\,0.73 \\
			NEACOs & 4 & 3 & 7.8 & 6.6 & 4.8 & 2.85 & 2.15 & 1.65 & 2.62\,|\,2.58 & 2.7\,|\,2.46 & 0.18\,|\,0.27 \\
			\hline
		\end{tabular}
		\label{Table:Slopes}
	\end{table*} 
	
	As for orbital inclination, C-complex group is divided into two subsets. The first one contains 23 objects with i\,$\leq$\,28$^\circ$, while the second group consists of 8 NEOs with i\,$\in$\,(39$^\circ$, 68$^\circ$). The largest inclination of $\sim$\,68$^\circ$ is shown by (5496) 1973 NA. 
	
	An anti-correlation between the visible spectral slope and T$_J$ has been reported by \citet{2008A&A...481..861L}, who studied the spectral properties of 41 asteroids having T$_J$\,<\,3. They noticed that the reddest of them have T$_J$\,<\,2.7. Generally, these bodies reside within unstable regions in the q vs.T$_J$ space, foreshadowing a recent injection onto such orbits. In our attempt to verify if our catalogue of low-T$_J$ NEOs describes a similar relation, we calculated the visible and NIR spectral slope and the corresponding errors for those objects with comet-like taxonomies. The slopes errors were calculated following the error propagation formula.
	
	The distribution of the 32 objects with known visible slopes is presented in Fig.~\ref{Fig:Vis_slopes}. In Table\,\ref{Table:Slopes} a summary of the slopes and T$_J$ calculations for various category of objects can be found. The median visible slope of all low-T$_J$ NEOs with cometary nature is S$_{vis}^{med}$\,=\,$4.15\,\pm\,3.9$\,per\,cent\,(1000\,$\si{\angstrom}$)$^{-1}$. This value is comparable with the one obtained by \citet{2008Icar..194..436D} of $3.7\,\pm\,2.9$\,per\,cent\,(1000\,$\si{\angstrom}$)$^{-1}$ (we note that this is a rough comparison because the computation reported by \cite{2008Icar..194..436D} uses the 0.44\,-\,0.92\,$\mu m$ spectral interval). 
	
	Because our sample set includes a small number of low-T$_J$ NEOs with known visible slope, we cannot highlight the anti-correlation between T$_J$ and the visible slope. With the exception of 3552 (in the lower-right part of Fig.~\ref{Fig:Vis_slopes}) and the Halley-type NEOs, 2007 VA85 (comet 333P - which is not plotted due to its low T$_J$ of 0.418), all NEOs are placed above T$_J$\,=\,2.5. Yet, a notable displacement exists between the D-group, having a median visible slope and T$_J$, (S$_{vis-D}^{med}$\,=\,$7\,\pm\,3.6$\,per\,cent\,(1000\,$\si{\angstrom}$)$^{-1}$, T$_{J}^{med}$\,=\,$2.65\,\pm\,0.73$) and the B\,/\,C-complex group with (S$_{vis-(B/C)}^{med}$\,=\,$-0.5\,\pm\,2$\,per\,cent\,(1000\,$\si{\angstrom}$)$^{-1}$, T$_{J}^{med}$\,=\,$3\,\pm\,0.15$). However, more data are needed to highlight this anti-correlation.
	
	The plot for the NIR slopes is presented in Fig.\,\ref{Fig:NIR_slopes}. The 51 objects for which we calculated the NIR slope show a median value of S$_{NIR}^{med}$\,=\,$1.46\,\pm\,1.72$\,per\,cent\,(1000\,$\si{\angstrom}$)$^{-1}$. However, we find objects which are widely distributed throughout the T$_J$ values interval (with a standard deviation of T$_{J}^{std}$\,=\,0.4) having both, high and small NIR slopes. Finally, the group of seven NEACOs show a median visible slope of S$_{vis}^{med}$\,=\,6.6\,$\pm$\,4.8\,\,per\,cent\,(1000\,$\si{\angstrom}$)$^{-1}$, slightly lower than of D-types group, because it also contains three objects with C\,/\,X-complex spectral types (thus, smaller visible slopes). Yet, we note that in both spectral regions, the NEACOs show redder surfaces compared to low-T$_J$ NEOs.

	\subsubsection*{The silicates, basaltic and miscellaneous spectral group}
	
	The silicate-like spectral group is the most abundant in the entire NEOs population. The Q\,/\,S-complex spectral types represent $\sim$\,50-60\,per\,cent \citep{2019A&A...627A.124P, 2019Icar..324...41B}. Our data shows that this ratio decreases substantially for NEOs with low-T$_J$ values (see Fig.~\ref{Fig:tax_all}). We report 71\,/\,150 (47.3\,per\,cent) NEOs with silicate-like spectral signatures. Most of them ($\sim$81\,per\,cent) are classified as S-complex, 21 S-, 19 Sr-, 12 Sq- and 6 Sv-type, while the fraction of Q-types is $\sim$13\,per\,cent (9 objects). Within this sample, the largest know NEO with D$_{eq}\,\sim$\,38\,km \citep{2011ApJ...743..156M}, namely (1036) Ganymed is included. It is characterized by a T$_{J}$\,=\,3.035, a perihelion q\,=\,1.24\,au and an S-type spectral signature \citep[e.g.][]{2004Icar..170..259B}. We also retrieved data of eight asteroids with end-member spectral types, one A-type (2019 HC), three R-types (394130, 466130, 2015 OL35) and four belonging to basaltic group (the V-types), namely 215188, 285944, 2001 EC and 2017 MB1. About 60\,per\,cent (42\,/\,71) of all silicate-like asteroids have available spectral information only in NIR range. Visible spectra were obtained for $\sim$\,25\,per\,cent (18\,/\,71) and a full VNIR spectrum have been recorded for 11 objects.  
	
	Compared to cometary-like objects, the most important features of silicate and basaltic asteroids reside in the NIR region. These spectral curves can be easily distinguished by the presence of the two absorption bands around 1 and 2\,$\mu m$, but their shape, depth and position are specific to each taxonomic type. The members of Q\,/\,S-complex group are associated with the most abundant meteorites group, ordinary chondrites \citep[e.g.][]{1984Icar...60...83G,1985Sci...229..160M,1996M&PS...31..699C}. This composition link has been directly confirmed by the sample returned Hayabusa mission from the S-type object Itokawa \citep{2011Sci...333.1113N}. Meanwhile, the objects spectrally classified as V-types show composition signatures similar to Howardite-Eucrite-Diogenite (HED) meteorites group \citep[e.g.][]{1977GeCoA..41.1271C}.
	
	It is generally accepted that a silicate-like composition cannot be associated with a cometary nature (i.e. having a considerably content of volatiles and organic material, which is specific to objects formed beyond the snowline). However, one notorious object which disputes this idea is C/2014 S3 (PANSTARRS). \citet{2016SciA....2E0038M} have reported that this small body has a long-period comet-like orbit, coming from the Oort cloud, but it is physically similar to an inner main belt rocky S-type asteroid. They detected a weak comet-like activity, which is consistent with volatiles sublimation and it was classified as a Manx object. They argue that during the planet-formation processes, C/2014 S3 was ejected from the inner solar system and its volatile content has been preserved for billions of years in the Oort cloud.
	
	The orbital elements of NEOs from the silicate spectral group show a widespread distribution. Their eccentricity ranges from 0.3 to 0.97, with the bulk of distribution ($\sim$72\,per\,cent) between 0.5 and 0.7. As for orbital inclination, most of them ($\sim$93\,per\,cent) have i\,$\leq$\,40$^\circ$, while four objects form a particular group of high-inclination bodies (i\,$\geq$\,57$^\circ$). The most noticeable case is (466130) 2012 FZ23, an R-type potentially hazardous asteroid (PHA) with i\,=\,75.3$^\circ$ (the highest value within the entire set, excepting 333P), T$_J$\,=\,2.39 and a Minimum Orbital Intersection Distance (MOID) with Earth of 0.02\,au. 
	
	Although 90\,per\,cent of all low-T$_J$ NEOs with Q\,/\,S-complex taxonomies analyzed in this work are on orbits with T$_J$\,$\geq$\,2.8, we found a particular group of four S-complex NEOs (see Table~\ref{S-extreme}), showing small values of T$_J$ and the lowest perihelia among entire catalogue. As a consequence of the low-perihelia and the high eccentricities (e\,$\geq$\,0.92), these NEOs are the subject of large temperature variations between perihelion and aphelion passages.  Below, we discuss the two most extreme cases.
	
		\begin{table}
		\centering
		\setlength\arrayrulewidth{0.5pt}
		\caption{Summary of all extreme cases of low-T$_J$ NEOs with small perihelion distances and with taxonomies from the silicate-like composition group. For each object, the following properties are shown: main designation, taxonomic type, Jovian Tisserand parameter (T$_J$), perihelion distance (q), orbital inclination (i), the surface temperature reached during the perihelion passage (T$_{per}$) and absolute magnitude (H).}
		\rowcolors[]{1}{gray!0}{gray!7}
		\begin{tabular}{ l c c c c c c}
			\hline
			\rowcolor{gray!30}
			Designation & Tax. & T$_J$ & q\,(au) & i\,($^\circ$)& T$_{per}$\,(K) & H\,(mag)\\
			\hline
			394130 & R & 2.3 & 0.082 & 33.4 & 960 & 17.1 \\
			465402 & Sr & 2.4 & 0.103 & 10.5 & 825 & 17.3 \\
			455426 & S & 2.6 & 0.197 & 6.83 & 610 & 18.6 \\
			331471 & Sv & 2.68 & 0.26 & 14.2 & 510 & 15.6 \\
			\hline
		\end{tabular}
		\label{S-extreme}
	\end{table}
	
	We found (394130) 2006 HY5 as the asteroid with the most extreme orbit in our sample, due to its perihelion of 0.082\,au, T$_J$\,=\,2.3 and the large surface temperature variation. Using Eq.\,\ref{Eq:temp}, we estimated a variation of $\Delta$T\,$\sim$\,840\,K between perihelion and aphelion, with a temperature reached during perihelion passage of $\sim$\,960\,K. Based on the albedo of p$_v$\,=\,0.157\,$\pm$\,0.071, the equivalent diameter (D$_{eq}$) is estimated to be $\sim$\,1.28\,km. The classification of its spectrum (retrieved from SMASS-MIT database) as an R-type has been derived based only on NIR observations. Although noisy at $\lambda$\,$\geq$\,1.5\,$\mu m$, the 1\,$\mu m$ band shows a good match with the R-type reference spectrum, but the 2\,$\mu m$ band is deeper. As of October 2020, this is the NEO with the lowest perihelion and a well-determined orbit (there are few others with smaller perihelia, but with uncertain orbits).
	
	\emph{(465402) 2008 HW1} has the second-lowest perihelion and T$_J$ among objects from this composition group. Its temperature variation is estimated to $\Delta$T\,$\sim$\,700\,K. We found a Sr-type classification for this asteroid reported in the SMASS-MIT database. The classification is derived only based on NIR data. Yet, some differences in the curve matching between Sr-type and object's spectrum seem to exist. The 1\,$\mu m$ feature is both deeper and steeper, while the 2\,$\mu m$ band display a broader shape. Also, a shallow feature can be observed between 2.2 and 2.5\,$\mu m$.

	Another particular case of asteroids with low-T$_J$ invariant is the presence of basaltic objects (corresponding to V-type). All four V-types included in our sample show large eccentricities (between 0.5 and 0.76), and their orbital inclinations are widespread. Those with sizes larger than 1 km show the highest inclinations 215188 (22.3$^\circ$, D\,$\sim$\,2.5\,km) and 285944 (53$^\circ$, D\,$\sim$\,1.5 km), while the low inclination V-types have sub-km sizes 2001 EC (0.6$^\circ$, D\,$\sim$\,0.4\,km) and 2017 MB1 (8.5$^\circ$, D\,$\sim$\,0.5 km). These are interesting objects because of their origin and orbital diversity. The main mechanism which led to the formation of basaltic asteroids is planetary differentiation. During this process, multiple layers with different densities form under an internal thermal gradient (reaching 1\,500\,$^\circ$C). In the early solar nebula, such temperatures had been maintained by heat production from radioactive decay of Al$^{26}$ and Fe$^{60}$ isotopes \citep[e.g.][]{2015aste.book..533S}. Hence, the iron-nickel layer sinks towards the centre, covered later by a silicate mantle and a basaltic crust \citep{2006mess.book..733M}.

	\section{DISCUSSIONS}
	\label{sec:discussions}
	
	The spectral catalogue presented in this work allows us to infer the number of low-T$_J$ NEOs with cometary nature. 
	Both spectral and dynamical features are taken into account for this analysis. Dynamically, a value of T$_J$\,$\leq$\,3.1 has been chosen as an upper condition, while being classified as a NEACOs object was used for the lower limit. As for physical features, we have considered the spectral types from B\,/\,C-complex (B, C, Cb, Cg, Ch, Cgh), the red slope and featureless spectra of D- and T-type and the X-complex (X, Xe, Xk, Xc, Xn) types (although for these X-complex bodies we need to know the albedo too, in order to confirm their nature) \citep{2002AJ....123.1039J, 2003A&A...398L..45L, 2005Icar..179..174A, 2006AJ....132.1346C, 2011Icar..212..180C, 2011Icar..216..309C,2012Icar..218..196D}.
	
	\begin{figure}
		\includegraphics[width = 8.8cm, height = 7.25cm]{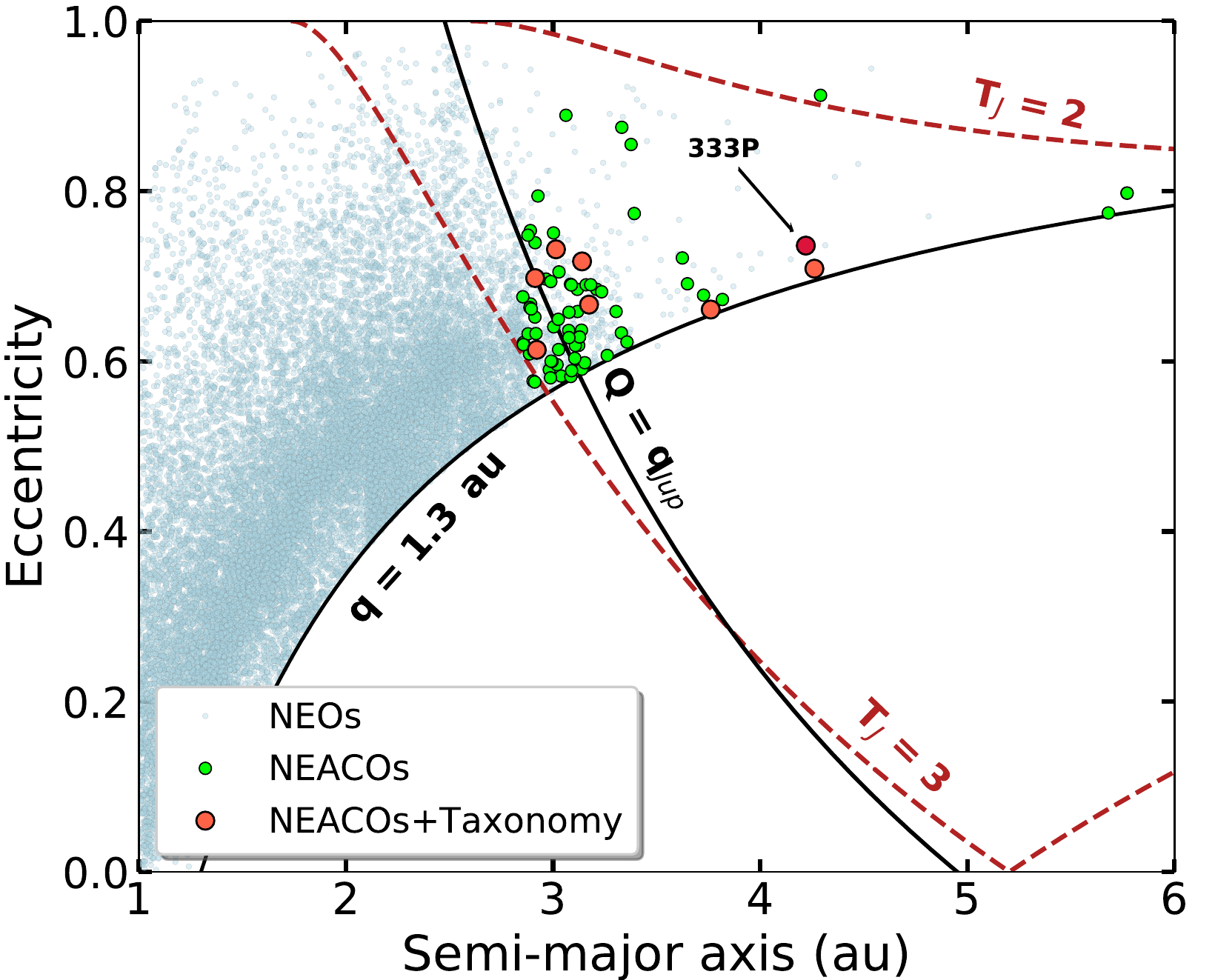}
		\caption{The orbital distribution in the \emph{e} vs. \emph{a} space of all NEACOs retrieved from Tancredi's list of ACOs (green dots) and of those NEACOs with known taxonomic classification (red dots). The entire population of NEOs (available on October 31, 2020) is plotted for reference (blue dots). The line of aphelion equals Jupiter's perihelion and the two limits of the Jovian Tisserand invariant (T$_J$\,=\,2 and 3, calculated for an orbital inclination i\,=\,0$^\circ$) are plotted for guidance. Four of the five  Halley-type NEACOs (T$_J \leq$ 2) are not shown in the figure because of their large semi-major axis.}
		\label{Fig:Tancredi_evsa}
	\end{figure}
	
	To identify the NEOs with a cometary origin, we sought to match the comet-like spectral types with dynamical properties. Firstly, we searched how many objects from Tancredi's list of ACOs are labelled as NEOs on cometary orbits (NEACOs). A total of 76 NEACOs were identified, out of which 71 asteroids show orbits similar to representatives from Jupiter Family Comets (T$_J\,\in$\,(2, 3.1)). The last five NEACOs, namely 1999 XS35, 2009 WN25, 2012 US136, 2014 PP69 and 2020 BZ12 display nearly-isotropic orbits, with T$_J$\,<\,2. \citet{2019MNRAS.487.2335I} had reported that 2009 WN25 has a primitive D-type composition and there is a possible link of this object with the November i-Draconids meteor shower. The distribution of these asteroids in the \emph{e} vs. \emph{a} space is shown in Fig.\,\ref{Fig:Tancredi_evsa} (green dots)
	
	Then, we checked how many NEACOs have been spectrally classified. Only seven objects (red dots - Fig.~\ref{Fig:Tancredi_evsa}) have a recorded spectrum, all of them with comet-like spectra, 3 D-type ((3552) Don Quixote, 85490, 2001 UU92), 1 X-type 248590, 1 C-type 475665, 1 T-type 485652 and 1 Xc-type 506437. The cometary nature of 248590 is also sustained by its albedo p$_v$\,=\,0.018\,$\pm$\,0.007 \citep{2011ApJ...743..156M}. This sample covers only objects with T$_J$ between (2, 3.1), which represents orbits similar to Jupiter Family Comets. Except 2007 VA85, later reclassified as the comet 333P\,/\,LINEAR, we did not observe any object with a Halley-type orbit (characterized by T$_J$\,<\,2). Although it has a T$_J$\,=\,0.418, 333P is placed in Fig.\,\ref{Fig:Tancredi_evsa} between T$_J$\,=\,2 and 3 curves (and next to (3552) Don Quixote). This happens because the figure is compiled for an orbital inclination i\,=\,0$^\circ$. In reality, the inclination difference between these two objects is $\sim$100$^\circ$. Thus, if the eccentricity would be represented as a function of both, semi-major axis and inclination, 333P would be placed above the surface of T$_J$\,=\,2, while 3552 would be placed below.
	
	As for orbital inclination, this group of NEACOs with known taxonomy includes four objects with i\,<\,10$^\circ$ and three with i larger than 20$^\circ$, with 248590 having the largest inclination of 52.3$^\circ$. Thus, we conclude that these seven NEACOs are dormant or extinct comets.  Also, it is important to note that no asteroid with other taxonomy than comet-like has been found in the list of NEACOs, which is consistent with the dynamical criterion of \citet{2014Icar..234...66T}.

	In the case of (3552) Don Quixote, the largest known D-type (D$_{eq}\,\sim$\,19\,km) within the NEOs population \citep{1987Icar...69...33H,2004Icar..170..259B}, our conclusion regarding its comet-like nature is in accordance with the cometary activity, detected in May 2018 by \citet{2020PSJ.....1...12M}. They observed Don Quixote during its perihelion passage in NIR, using the Spitzer Space Telescope and in visible with the SOAR and Magellan ground-based telescopes. Besides the NIR emissions (4.6\,$\mu m$) of molecular transitions from CO$_2$ or CO molecules, which were also detected during the previous passage \citep{2014ApJ...781...25M}, they observed for the first time a cometary activity in visible wavelengths. These optical signatures are associated with the light reflected by ejected dust particles from its coma. They concluded that 3552 is a weakly-active comet. The surface albedo of 0.03 \citep{2014ApJ...781...25M} is in agreement with its nature.
	
	The spectral observations show that NEOs with a comet-like composition represent a higher fraction in the sub-population of low-T$_J$ NEOs and they become dominant once T$_J$ decreases below $\sim$\,2.8. In the following section, we will attempt to make a bias correction of this distribution, by considering both the absolute magnitude and the Tisserand parameter. Then, we will compare our results with other previous studies. Last but not least, we will bring into discussion the relation between the spectral classification and the rotational properties of these bodies.

	\subsection{Bias-corrected taxonomic distribution of low-T$_J$ NEOs}
	
	Because of their eccentric orbits and low-albedo surfaces, the asteroids with a cometary origin have a short time interval during which they can be observed, using ground-based telescopes. Moreover, this visibility window repeats few times per century. Thus, both the detection and spectral characterization opportunities are influenced by these factors, and their currently known number is much lower compared with the rest of NEOs. A hint regarding the discovery completeness is shown in Fig.\,\ref{Fig:CumHNEA}. The linear trend (when Oy axis is in logarithmic scale) described by the curves from Fig.\,\ref{Fig:CumHNEA} suggests that the large majority of objects with an absolute magnitude brighter than $\sim$19\,mag, were discovered \citep{2015Icar..257..302H}.

	The attempt to remove the selection effects from the distribution of the observed asteroids has been performed by many authors, including \citep{1971NASSP.267...51C, 1971NASSP.267..187K, 1982Sci...216.1405G, 1999PhDT........50B, 2003Icar..162...10M, 2004Icar..170..295S, 2013Icar..226..723D}. Our approach for removing the observational biases follows a similar methodology as these references. We have broken up the sample of low-T$_J$ NEOs into an absolute magnitude grid, with a bin width of 1\,mag. Then, we divided the sample into two parts with respect to the Tisserand parameter, by using the T$_J$\,=\,2.8 as a threshold. Due to the low number of objects at bright and faint absolute magnitudes, we limited the taxonomic distribution for magnitudes between 14 and 21\,mag.
	
	\begin{equation}
	N^{i , (tax)}_{\footnotesize {unbiased}} = \frac{N^{i , (tax)}_{\footnotesize {obs}}}{N^{i , (all)}_{\footnotesize {obs}}} \boldsymbol{\cdot} N^i_{\footnotesize {discovered}}
	\label{eq:debias}
	\end{equation}
	
	We compute in each bin the fraction of each spectral group, relative to the number of asteroids with known taxonomy. After that, each fraction is multiplied by the total number of discovered NEOs in that bin. The  Eq.~\ref{eq:debias} summarizes this computation. In this equation, \emph{i} is the bin number, $N^{i , (tax)}_{unbiased}$ is the unbiased number of asteroids from the i-th bin, $N^{i , (tax)}_{obs}$ is the observed number of asteroids with taxonomies from each composition group from the i-th bin, $N^{i , (all)}_{obs}$ is the total number of asteroids with known taxonomy from the i-th bin and $N^i_{discovered}$ is the total number of discovered asteroids, having T$_J$\,$\leq$\,3.1 from the i-th bin.
	
	\begin{figure}
		\centering 
		\includegraphics[width = 9.5cm]{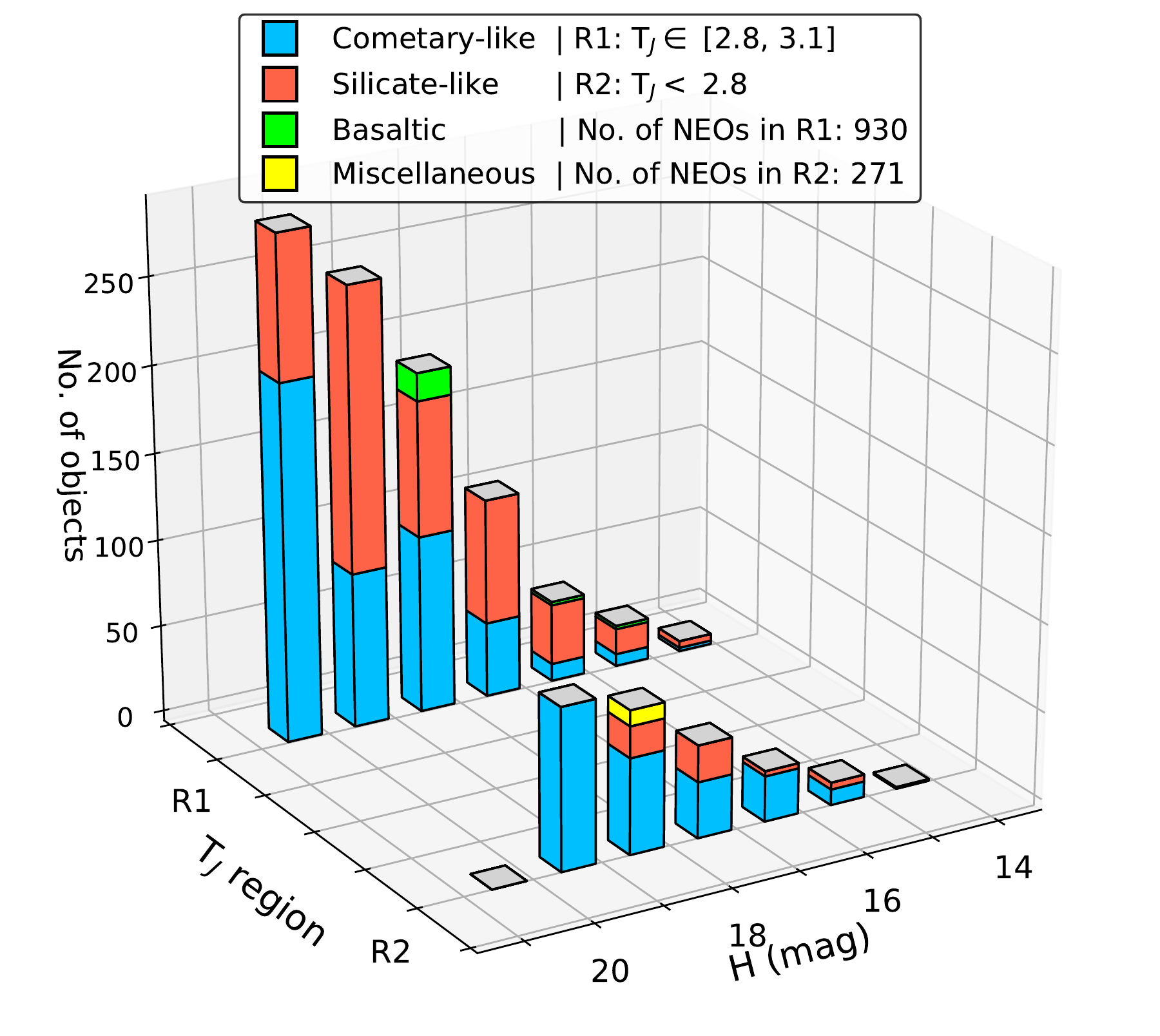}
		\caption{The bias-free taxonomic distribution of low-T$_J$ NEOs, computed for objects with absolute magnitudes H between 14 and 21\,mag. A bin width of 1\,mag has been used. We have considered two possible regions (R1 and R2) for the Jovian Tisserand parameter, delimited by T$_J$\,=\,2.8. The choice of this boundary-value has been motivated by the inversion of composition ratio between the cometary and silicate-like asteroids, which takes place around T$_J$\,=\,2.8 in the observed taxonomic distribution from Fig.\,\ref{Fig:tax_all}.}
		\label{Fig:Debias}
	\end{figure}
	
	The resulted unbiased taxonomic distribution is illustrated in Fig.~\ref{Fig:Debias}. This figure highlights that the fraction of objects with a possible cometary origin is varying with the absolute magnitude and with T$_J$ parameter. With our limited statistics, we predict that the objects with cometary-like spectra can represent 49.4\,per\,cent (459 out of 930) for T$_J$\,$\in$\,(2.8, 3.1) and 79.7\,per\,cent (216 out of 271) for T$_J$\,$\leq$\,2.8. In both cases, this ratio shows an increasing trend toward faint absolute magnitudes (small objects). This suggests  an upper limit for the objects with a cometary nature of (10.4\,$\pm$\,2.2)\,per\,cent from the total NEOs population with absolute magnitude H\,$\in$\,(14, 21)\,mag. However, this limit includes objects which may originate in the main asteroid belt and had arrived on these orbits through various orbital resonances.

	Opposite, in order to compute the lower limit of NEOs with a cometary origin, we took into account that all NEACOs have spectra compatible with a comet-like nature. However, the criterion of \cite{2014Icar..234...66T} did not consider the objects with uncertain orbits. The Minor Planet Center database provides an orbital uncertainty parameter (namely U-code), ranging from 0 to 9. Within \cite{2014Icar..234...66T} criterion, a sufficient precise orbit is considered when U-code\,$\leq$\,2. If the object has U-code\,>\,2, it will be rejected from the selection procedure. Thus, we need to take into account the possible contribution of those objects with uncertain orbits, and which may be confirmed as NEACOs by further surveys. 
	
	To estimate this contribution, we calculated what is the fraction of NEACOs among the low-T$_J$ NEOs with well-known orbits (satisfying the U-code\,$\leq$\,2). Out of all 1\,299 NEOs with T$_J$\,$\leq$\,3.1 and H\,$\in$\,(14, 21)\,mag, 783 (60.3\,per\,cent) have U-code\,$\leq$\,2 and 516 (39.7\,per\,cent) have U-code\,>\,2. Thus, the NEACOs with well-known orbits represent 8.4\,per\,cent (66\,/\,783) from the low-T$_J$ NEOs sample with certain orbits.

	Now, we multiply this fraction with the number of low-T$_J$ NEOs having U-code\,>\,2, to estimate the additional contribution; 0.084\,$\cdot$\,516 = 43. Finally, the fraction of inactive or extinct comets is obtained by dividing the number of NEACOs having H\,$\in$\,(14, 21)\,mag to the number of NEOs, within the same magnitude interval and multiplying with the fraction of comet-like objects in the NEACOs with known taxonomy sample (Eq.~\ref{lowerlim}). We obtain a value of (1.5\,$\pm$\,0.15)\,per\,cent for the lower limit of NEOs with cometary origin. 
	
	\begin{equation}
	\frac{66 + 43}{7026} \boldsymbol{\cdot} 100\% = (1.5 \pm 0.15)\% 
	\label{lowerlim}
	\end{equation}
	
	The assumption of a cometary origin for the asteroids satisfying \citet{2014Icar..234...66T} criterion is supported also by the study of \citet{2018A&A...618A.170L}. They analyzed the spectra of 29 ACOs (most of them from the main asteroid belt), and found that all (except one object with low SNR) share spectral features with comets and all ACOs in long-period orbits (known as Damocloids) have D-type taxonomy, whereas in the JFCs-ACOs group 60\,per\,cent were D-type and 40\,per\,cent were X-type. They concluded that all objects from Tancredi's list are inactive comets. Within our short list of seven NEACOs with known taxonomy, we report a similar dominance of D-types (3\,/\,7). 
	
	\subsection{Comparison with previous studies}
	
	The simulations performed by \citet{2014Icar..238....1F, 2015P&SS..118...14F} for 139 NEOs (2\,<\,T$_J$\,<\,3 and Q\,>\,4.8\,au) and 58 JFCs (2\,<\,T$_J$\,<\,3 and q\,<\,1.3\,au) had shown that most NEOs have stable orbits on a time-scale of $\sim$\,10$^4$\,yr, while the majority of JFCs are highly unstable due to frequent close-encounters with Jupiter. Their capture time-scale into near-Earth orbits is less than a few 10$^3$\,yr. Nevertheless, they found a small group of nine NEOs moving on cometary orbits, similar to JFCs and foreshadowed that these objects might be inactive comets. One of them (2003 WY25) has been associated with the active comet 289P/Blanpain \citep{2006AJ....131.2327J}. Another object from this small group of NEOs, (85490) 1997 SE5, is also one of the seven NEACOs with a cometary-like spectrum  analyzed in this paper. Supposing that all nine NEOs are of cometary origin, they proposed a fraction of 17\,per\,cent for the ratio of inactive to active comets within the NEOs population with Q\,>\,4.8\,au.
	
	Our findings regarding an increased abundance of objects with a comet-like composition in the low-T$_J$ region are in agreement with other studies \citep{2008Icar..194..436D, 2008A&A...481..861L}. Our inferred limits (1.5\,$\pm$\,0.15) and (10.4\,$\pm$\,2.2)\,per\,cent for the objects with cometary nature in the NEOs population (with an absolute magnitude limit brighter than 21\,mag) are of the same order as the result of \citet{2008Icar..194..436D}, who report 8\,$\pm$\,5\,per\,cent, using a catalogue of 55 asteroids with T$_J$\,<\,3 and with known taxonomy. However, we found that this ratio varies with T$_J$ and with the absolute magnitude.
	
	Thus, T$_J$\,$\sim$\,3 can act not only as a dynamical border, but as a composition limit too. We found that the comet-like objects are the dominant group only for T$_J$\,$\leq$\,2.8. Between 2.8 and 3.1 there is a mixed population of objects, because this region marks the transition between the population with T$_J$\,>\,3 ($\sim$\,90\,per\,cent of NEOs population), dominated by Q\,/\,S-complex objects and the low-T$_J$ NEOs region, with a higher abundance of comet-like material. The dominance of Q\,/\,S-complex asteroids in the NEOs population is explained by a continuous replenishing with material mainly from the inner main belt (IMB), which is dominated by silicate bodies \citep{2014Natur.505..629D, 2016Icar..268..340C,2019Icar..324...41B}.

	\citet{2019Icar..324...41B} concluded that the low-albedo bodies show a dominated escape region from the outer main belt (OMB), with a strong contribution of D-types from the JFCs and 2:1 MMR. However, because the JFCs and 2:1 MMR regions have a very low probability for delivering asteroids to the overall NEOs population, D-type objects represent a lower fraction. Most of them are transported onto comet-like orbits with T$_J$\,<\,3, in agreement with the 16 D-type asteroids within our spectral catalogue. Excepting the largest of them, (3552) Don Quixote, their size ranges from $\sim$\,0.3 to 7\,km. Within our previous study, \citet{2019A&A...627A.124P} studied the spectral data of 76 NEOs with T$_{J}$\,>\,3 and sizes ranging from 0.25 to 5.5\,km and they did not find any D-type asteroid.
	
	Another possible source of low-albedo NEOs is represented by several families in the inner main belt. These are Polana, Erigone, Sulamitis, Clarisa, Klio, Chaldaee, Chimaera, and Svea. A significant fraction of asteroids belonging to these families has been spectrally characterized by the \emph{PRIMASS} survey \citep{2016A&A...586A.129M,2018A&A...610A..25M,2019A&A...630A.141M, 2016Icar..266...57D, 2016Icar..274..231P}. Most members of these collision families have been classified as C-, X-complex, and B-, T-type objects. Thus, we cannot exclude that the upper limit we found for asteroids with comet-like nature is overestimated due to objects originating from these families.
	
	Within the set of NEOs studied in this paper, we reported a group of four low-perihelion asteroids, with taxonomies falling into S-complex. While at q\,$\geq$\,0.5\,au the abundance of comet-like and silicates objects is roughly identical, towards lower q the number of low-albedo asteroids starts to decrease and finally, for q\,$\leq$\,0.25\,au, a lack of B\,/\,C-complex, D-, T-, X-types NEOs exists. The asteroid (170502) 2003 WM7 has the lowest perihelion (q\,$\sim$\,0.3\,au) among NEOs with comet-like composition, but higher than the objects listed in Table\,\ref{S-extreme}. This lack of low-albedo NEOs is predicted by the results of \citet{2016Natur.530..303G}. Based on albedo values, they concluded that the observed deficit of primitive NEOs in the Sun's proximity is a consequence of super-catastrophic breakups experienced by these objects, when they come at few solar radii far from the Sun. Generally, primitive asteroids are more fragile and mechanisms such as large temperature variations, thermal fragmentation or erosion by radiation pressure have a more intense overall effect upon some low-albedo bodies, than on silicate-like objects. Notice that the predictions of \citet{2016Natur.530..303G} imply a critical perihelion of q$_*$\,=\,(0.076\,$\pm$\,0.0025)\,au, below which catastrophic breakups happen. None of the low-T$_J$ NEOs within our set has q below this critical value. Thus, up to this point, the low-perihelion objects we retrieved are in agreement with the predicted disruption distance q$_*$.

	Another interesting feature is that even though at larger perihelia (q\,$\geq$\,0.3\,au) the S-, Q- and the intermediate Sq-type objects are evenly distributed relative to q, there is a peak of S-complex asteroids among very low-perihelion NEOs. According to laboratory experiments on ordinary and carbonaceous meteorites, \citet{2014Natur.508..233D} had emphasized that thermal fragmentation of centimetre-sized rocks is the dominant mechanism in regolith production on the surface of small asteroids and is several orders of magnitude faster than the regolith production by meteoroids impact. This fragmentation is controlled by cyclic diurnal temperature variations. Hence, the abundance of NEOs with surfaces unaffected by space weathering processes (resembling Q-type spectra) should increase toward lower perihelia, since diurnal temperatures are higher. 
	This is in agreement with the findings of \citet{2019A&A...627A.124P}, who studied the spectral properties of NEOs with T$_J$\,$\geq$\,3, and reported an abundance of Q-type NEOs at lower q. Within their catalogue of 76 NEOs, they found that all asteroids with perihelia between $\sim$\,0.2 and 0.6\,au show Q-type spectra and diameters from $\sim$0.5 to 1.5\,km. Based on a set of 195 NEOs, classified as Q-, Sq- or S-type, \citet{2019Icar..324...41B} also reported a similar result.

	Our observations indicate that low-perihelia NEOs with T$_J$\,$\leq$\,3.1 show surfaces strongly affected by space weather and meteorites bombardment (similar to S-complex taxonomy). However, since the four extreme NEOs reported in this paper have a mean perihelion ($\overline{q}$\,=\,0.16\,au) smaller than of NEOs with the lowest q analyzed by \citet{2019A&A...627A.124P} and \citet{2019Icar..324...41B}, a direct comparison is not appropriate. A possible explanation of the observed Q-type deficit at very small q is that these asteroids display \emph{saturated} surfaces, as a consequence of more energetic solar wind and meteorites flux. \citet{2019Icar..324...41B} have emphasized that this saturation state may be reached after multiple episodes of resurfacing, followed by long periods of weathering. Once the surface of dust grains have been altered on all sides, any new resurfacing episode will not expose other grains with \emph{fresh} sides.

	\subsection{Taxonomy vs. rotational properties}
	
	To verify how the spectral types and rotational properties are correlated, we searched the spin periods and light-curve amplitudes from the LCDB database. Information for a total of 83 NEOs has been retrieved. Almost 46\,per\,cent (38\,/\,83) are part of the silicate-like spectra group, 50\,per\,cent (42\,/\,83) display comet-like features and only three are part of the basaltic group.
	
	\citet{1996LPI....27..493H} studied the rotational properties of 688 bodies and concluded that most asteroids have a fragile internal structure and they will not survive against centrifugal acceleration, if the spin period decreases below 2\,h (or equivalent to 12\,rot\,d$^{-1}$). Other studies \citep[e.g.][]{2000Icar..148...12P,2009Icar..202..134W, 2015AJ....150...75W} have reported similar results about potentially breakups for asteroids with equivalent diameters greater than $\sim$\,300\,m. Also, more recent studies on (101955) Bennu \citep{2019NatAs...3..341D} and (162173) Ryugu \citep{2019Sci...364..268W} show a rubble-pile structure (i.e. collection of boulders with different sizes, separated by voids) of these bodies and it is very likely that many other asteroids have a similar internal structure. The 12\,rot\,d$^{-1}$ spin limit is known in literature as the \emph{rubble-pile} spin barrier.
	
\begin{figure}
		\centering
		\includegraphics[width=8.75cm]{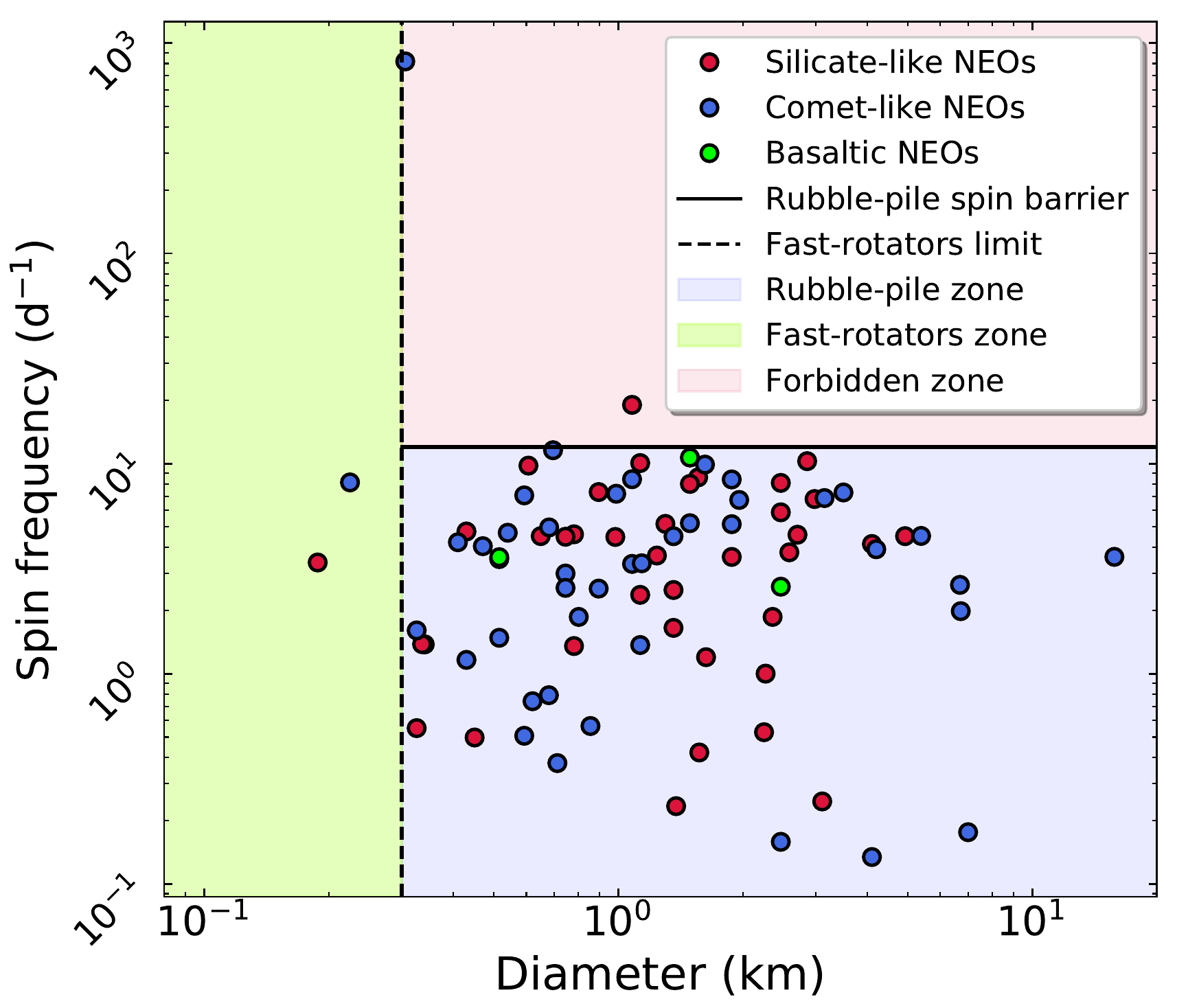}
		\caption{The distribution of objects with known spectral classification in the spin frequency vs. equivalent diameters space. Both quantities have been retrieved from literature, by accessing the Asteroids Light-Curve Database from the Minor Planet Center. We found this information for 83 NEOs. The \emph{rubble-pile} spin barrier has been chosen to 12\,rot d$^{-1}$, but only for those bodies with sizes larger than $\sim$300\,m. This value was proposed by \citet{1996LPI....27..493H}.}
		\label{Fig:FreqD}
	\end{figure}

	In Fig.~\ref{Fig:FreqD} we have plotted the log-log distribution of spin frequencies (in units of rot\,d$^{-1}$) relative to the object's equivalent diameter. Three zones are distinguishable: the \emph{rubble-pile zone} (blue area), the \emph{fast-rotators zone} (green area) and the \emph{forbidden zone} (red area), delimited by two borders. Most objects (79\,/\,83) reside in the \emph{rubble-pile zone}, in agreement with the evidences of a low number of extreme cases. Here, the cometary-like objects are more abundant in the sub-km zone  and they also display the smallest values of spin rate: 0.13\,rot\,d$^{-1}$ (16064), 0.16\,rot\,d$^{-1}$ (163732) and 0.18\,rot\,d$^{-1}$ (53319), all of them being part of C-group. In the extreme right part, we find the largest known D-type NEO, (3552) Don Quixote (D$_{eq}\,\sim$\,19\,km), with a period of 6.6\,h.
	
	For equivalent diameters, less than 300\,m, an object can rotate faster than 12\,rot\,d$^{-1}$, without being disrupted. Here we find the \emph{fast-rotators zone}, where gravity is no longer the main mechanism responsible for avoiding an object to breakup. Only 2\,/\,83 bodies had been found with such properties: 2013 SW24 (D$_{eq}$\,$\sim$\,190\,m, 3.4\,rot\,d$^{-1}$) and 2015 CA1 (D$_{eq}$\,$\sim$\,220\,m, 8.1\,rot\,d$^{-1}$). The strong cohesion of these representatives could be explained through a structure of small grains, trapped in a boulders mixture. Several studies  \citep[e.g.][]{2010Icar..210..968S,2014M&PS...49..788S, 2015Icar..247....1S} showed that attraction through van der Walls forces and friction between regolith grains can increase the internal resistance, allowing super-fast rotations. Also, the findings of \citet{2010Icar..210..968S} indicate that in the case of small objects, the van der Walls forces may represent a mechanism equally important as the gravity, for preventing further breakups.
	
	The red zone has been assigned as the \emph{forbidden zone}, where we find objects with D$_{eq}$\,>\,300\,m experiencing rotations unexplained by a self-gravity based structure. The increasing or decreasing of spin velocity can be very efficiently induced by the YORP spin-up / down, especially for objects with equivalent diameters less than 10\,km \citep{2000Icar..148....2R, 2015aste.book..509V}. Depending on torque orientation, relative to spin direction, an object can experience acceleration or deceleration \citep{2013MNRAS.430.1376R}, which could lead to a pass above the \emph{rubble-pile} spin barrier. Only a small number of such extreme cases are known in literature: 2001 OE84 \citep{2002ESASP.500..743P} and 2000 GD65 \citep{2015DPS....4740202P} among them. In this region, we found two objects, 2019 HC and 2001 UC5. 
	
	\emph{2019 HC} is the only pure olivine A-type asteroid retrieved from literature. It has an equivalent diameter of $\sim$\,1\,km and a spin frequency of 19\,rot d$^{-1}$. If only gravity would act as the principal mechanism against centrifugal breakup, its density should be $\sim$\,7.3\,g\,cm$^{-3}$. This value is even greater than most iron meteorites \citep{2012P&SS...73...98C}. Consequently, it is clear that other mechanisms such as strong cohesion or a monolithic structure would be more suitable to explain this extreme case.
	
	\emph{2001 UC5} has a D-type taxonomy recorded in the SMASS-MIT database, derived based on visible data. With a spin frequency of $\sim$\,820\,rot\,d$^{-1}$ and D$_{eq}$\,$\sim$\,300\,m, it is the most extreme case we have found. Yet, we investigated its spectra using the M4AST tool and it seems to be more appropriate to an X-complex type, rather than D-type. Objects with D-type taxonomy are not characterized by a strong internal resistance at all, which is more than necessary to such case. However, there is a large uncertainty about its albedo, and consequently about its size.

	\section{CONCLUSIONS}
	\label{sec:conclusions}
	
	Here we summarize the results of our work. In this paper, we have analyzed the spectral properties of 150 NEOs with low-Tisserand invariant (T$_{J}$\,$\leq$\,3.1), in order to estimate their possible cometary nature. Our main conclusions are:
	
	\begin{enumerate}[1)]
		
		\item The compositional map described by the near-Earth objects with low-T$_{J}$ differs from the equivalent one of the entire NEOs population. According to the observational bias-free taxonomic distribution, computed in this work for absolute magnitudes H\,$\in$\,(14, 21)\,mag, we predict that the asteroids with spectra similar to cometary nuclei represent the dominant composition group (56.2\,per\,cent) in the T$_J\,\leq$\,3.1 region, the silicate-like objects (which are dominant in the NEOs population) count only for 41.4\,per\,cent and the basaltic objects represent $\sim$1.6\,per\,cent.
		
		\item The composition ratio between the cometary and silicate objects varies significantly with respect to T$_{J}$. An inversion of this ratio can be seen around T$_J$\,=\,2.8. Between T$_J$\,$\in$\,(2.8, 3.1) there is a mixed population, with the primitive and silicate objects having comparable abundances (49.3 and 48.5\,per\,cent, respectively). As for T$_J$\,$\leq$\,2.8, the comet-like objects have the majority of 79.7\,per\,cent. Thus, the T$_J$ invariant can act not only as a dynamical border, but as a composition limit too. Also, we found that these abundances vary with the absolute magnitude. They show an increasing trend of the cometary objects' fraction toward fainter absolute magnitudes (thus, toward smaller sizes).
		
		\item Our sample of primitive asteroids describe a moderate anti-correlation between visible slope and T$_J$ (which has been reported in other studies). A displacement exists between the B\,/\,C-complex group with S$_{vis-(B/C)}^{med}$\,=\,$-0.5\,\pm\,2$\,per\,cent\,(1000\,$\si{\angstrom}$)$^{-1}$, T$_{J}^{med}$\,=\,$3 \pm 0.15$ and D-types group, having S$_{vis-D}^{med}$\,=\,$7\,\pm\,3.6$\,per\,cent\,(1000\,$\si{\angstrom}$)$^{-1}$, but T$_{J}^{med}$\,=\,$2.65\,\pm\,0.73$. Yet, an improvement of this statistic is needed, in order to highlight a stronger dependence.
		
		\item Even though the D-type asteroids represent only 3\,per\,cent within the NEOs population and are rare for T$_J$\,>\,3, we retrieved a total of 16 objects in the T$_J$\,$\leq$\,3.1 regime. This is in agreement with the high source region probability of JFCs and 2:1 MMR to inject D-types objects onto comet-like orbits (T$_J$\,$\leq$\,3.1) in the near-Earth space \citep{2019Icar..324...41B}.
		
		\item Within the analyzed catalogue of spectral data, we found a particular group of S-complex asteroids, showing low-T$_J$ values (2.3 to 2.7) and the lowest perihelia (q\,<\,0.3\,au) among the entire list of objects. The observed dominance of silicate asteroids at low perihelia is in agreement with the predictions that low-albedo bodies, with a primitive composition, have a fragile internal structure and breakup to higher perihelia. Nevertheless, all studied low-T$_J$ objects show perihelia larger than q$_*$\,=\,(0.076\,$\pm$\,0.0025)\,au, a limit proposed by other previous studies below which thermal catastrophic breakups take place.
		
		\item The majority of objects studied in this paper present a low internal strength. Considering the distribution of spin frequencies, with respect to their sizes,  we found the bulk of asteroids residing between the \emph{rubble-pile} spin barrier (12\,rot\,d$^{-1}$) and the \emph{fast-rotators} spin barrier (D\,$\sim$\,300\,m). We found only two bodies with equivalent diameters less than $\sim$\,300\,m, but both of them with spin rates less than 12\,rot\,d$^{-1}$. Nevertheless, we report two extreme cases with equivalent diameters larger than 300\,m and spin rates greater than 12\,rot\,d$^{-1}$, namely 2019 HC and 2001 UC5. 
		
		\item In our attempt to compute the fraction of NEOs with a comet-like nature and H\,$\in$\,(14, 21)\,mag, we have considered the spectra similar to comets nuclei (and\,/\,or albedo p$_v$\,<\,0.12 if it was available) as for physical features and two dynamical limits. The upper limit was T$_J$\,$\leq$\,3.1 and the lower one was that the object must be classified as a NEACOs, according to \citet{2014Icar..234...66T} criterion. Considering these aspects, we conclude that the fraction of NEOs with a cometary nature and H\,$\in$\,(14, 21)\,mag varies between (1.5\,$\pm$\,0.15) and (10.4\,$\pm$\,2.2)\,per\,cent. However, we note that the upper limit may also include bodies without a comet-like nature.
		
	\end{enumerate}

	\section*{Data availability}
	The data underlying this article will be shared on reasonable request to the corresponding author. We intend to upload the data to a public repository accessible trough Virtual Observatory tools.
	
	\section*{Acknowledgments}
	
	The asteroid spectra were acquired using Isaac Newton Telescope, NASA Infrared Telescope Facility as well as the AIRA ROC remote facility. The authors acknowledges the Spanish and Dutch allocation committees for the following observing runs, which allowed the data acquisition: N8/2014A, C97/2015A, N4/2015A, C26/2015B and N2/2015B. In addition, other data was acquired during some ING S/D nights (service time) and a few Spanish CAT service nights. We thank all the telescope operators and observers for their contribution. 
	Part of the data used in this publication was obtained and made available by the the MIT-UH-IRTF Joint Campaign for NEO Reconnaissance. The IRTF is operated by the University of Hawaii under Cooperative Agreement no. NCC 5-538 with the National Aeronautics and Space Administration, Office of Space Science, Planetary Astronomy Program. The MIT component of this work is supported by NASA grant 09-NEOO009-0001, and by the National Science Foundation under Grants Nos. 0506716 and 0907766. Any opinions, findings, and conclusions or recommendations expressed in this material are those of the authors and do not necessarily reflect the views of MIT-UH-IRTF Joint Campaign.
	The work of GS, MP, and RMG was supported by a grant from the Romanian National Authority for  Scientific  Research -- UEFISCDI, project number PN-III-P1-1.1-TE-2019-1504. This work was developed in the framework of EURONEAR collaboration. MP, JL, JdL acknowledge support from the ESA P3NEOI and NEOROCKS projects.
	The paper make use of data published by the following web-sites Minor Planet Center\footnote{\url{https://www.minorplanetcenter.net/}}, JPL Small-Body Database Browser \footnote{\url{https://ssd.jpl.nasa.gov/sbdb.cgi}}, and SMASS - Planetary Spectroscopy at MIT \footnote{\url{http://smass.mit.edu/}}. We thank the anonymous reviewer for the insightful comments and suggestions which helped us to improve our paper.

	%%%%%%%%%%%%%%%%%%%%%%%%%%%%%%%%%%%%%%%%%%%%%%%%%%
	
	
	
	\bibliographystyle{mnras}
	\bibliography{Source_files.bib} % if your bibtex file is called example.bib

	% \begin{thebibliography}{99}
	% \bibitem[\protect\citeauthoryear{Author}{2012}]{Author2012}
	% Author A.~N., 2013, Journal of Improbable Astronomy, 1, 1
	% \bibitem[\protect\citeauthoryear{Others}{2013}]{Others2013}
	% Others S., 2012, Journal of Interesting Stuff, 17, 198
	% \end{thebibliography}
	
	%%%%%%%%%%%%%%%%%%%%%%%%%%%%%%%%%%%%%%%%%%%%%%%%%%
	
	%%%%%%%%%%%%%%%%% APPENDICES %%%%%%%%%%%%%%%%%%%%%
	%\clearpage
	\appendix
	\clearpage
	%\newpage
	\onecolumn
	\section{The table of low-T$_J$ NEOs with known spectral data}
	\LTcapwidth=\textwidth
	\rowcolors[]{1}{gray!0}{gray!7}
	\begin{longtable}{l c c c c c c c c c c c c}
		\caption{The catalogue of all 150 low-T$_J$ NEOs {for which spectral data are available}. Here, their main orbital and  physical features can be found: asteroid's designation, Jovian Tisserand invariant, perihelion, aphelion, semi-major axis, eccentricity, orbital inclination, minimum orbital intersection distance with respect to Jupiter, taxonomic classification  assigned in this study, absolute magnitude, albedo, equivalent diameter and the spin period. On the \textit{p$_v$} column, the albedo values marked with * are the ones retrieved from the literature. For the rest of objects (when the albedo was not available), we have used the average albedo of each spectral type.}
		\label{neacos_table} \\
		\hline
		\rowcolor{gray!30}
		\textbf{Designation} & \textbf{T$_J$} & \textbf{q\,(au)} & \textbf{Q\,(au)} & \textbf{a\,(au)} & \textbf\,{e} & \textbf{i\,($^\circ$)} & \textbf{MOID\,(au)} & \textbf{Tax.} & \textbf{H\,(mag)} & \textbf{pv} & \textbf{D$_{eq}$\,(km)} & \textbf{P\,(h)}\\\hline
		\endfirsthead
		\multicolumn{12}{c}{Table 1: continuing}\\\hline
		\rowcolor{gray!30}
		\textbf{Designation} & \textbf{T$_J$} & \textbf{q\,(au)} & \textbf{Q\,(au)} & \textbf{a\,(au)} & \textbf\,{e} & \textbf{i\,($^\circ$)} & \textbf{MOID\,(au)} & \textbf{Tax.} & \textbf{H\,(mag)} & \textbf{pv} & \textbf{D$_{eq}$\,(km)} & \textbf{P\,(h)}\\\hline
		\endhead
		1036 & 3.03 & 1.244 & 4.09 & 1.87 & 0.318 & 26.7 & 1.946 & S & 9.2 & 0.238\textbf{*} & 37.67 & 10.30 \\ 
		1580 & 3.08 & 1.127 & 3.27 & 2.2 & 0.487 & 52.1 & 1.86 & B & 14.7 & 0.08\textbf{*} & 4.2 & 6.13 \\ 
		3552 & 2.33 & 1.24 & 7.28 & 4.26 & 0.709 & 31.1 & 0.44 & D & 13 & 0.03\textbf{*} & 15.79 & 6.67 \\ 
		4015 & 3.08 & 0.968 & 4.28 & 2.63 & 0.631 & 2.8 & 1.22 & B & 16 & 0.05\textbf{*} & 1.96 & 3.57 \\ 
		4197 & 3.09 & 0.522 & 4.07 & 2.29 & 0.773 & 12.6 & 1.27 & Sq & 14.9 & 0.239\textbf{*} & 2.98 & 3.54 \\ 
		4401 & 3.07 & 1.125 & 4.04 & 2.58 & 0.564 & 26.6 & 2.24 & Sr & 16 & 0.227 & 1.88 & 6.67 \\ 
		5496 & 2.55 & 0.887 & 3.98 & 2.43 & 0.636 & 68 & 2.67 & C & 16 & 0.05 & 1.88 & 2.86 \\ 
		6178 & 3.04 & 1.178 & 4.46 & 2.82 & 0.582 & 4.3 & 0.522 & X & 15.8 & 0.15\textbf{*} & 3.15 & 3.50 \\ 
		7092 & 3.01 & 0.767 & 4.3 & 2.54 & 0.698 & 17.8 & 1.51 & C & 15 & 0.05 & 2.84 & - \\ 
		7888 & 3.06 & 0.818 & 4.05 & 2.44 & 0.664 & 26.1 & 1.46 & S & 15.1 & 0.257\textbf{*} & 2.86 & 2.34 \\ 
		8201 & 3.03 & 0.744 & 4.33 & 2.54 & 0.707 & 9.6 & 0.661 & Q & 15.7 & 0.154\textbf{*} & 2.27 & 23.95 \\ 
		9400 & 2.96 & 1.087 & 4.09 & 2.59 & 0.58 & 36 & 2.48 & S & 14.8 & 0.211 & 3.11 & 97.10 \\ 
		10150 & 3 & 1.084 & 3.67 & 2.38 & 0.544 & 46 & 2.44 & Sr & 15.4 & 0.266 & 2.47 & 2.97 \\ 
		16064 & 3.02 & 1.171 & 4.53 & 2.85 & 0.589 & 4.5 & 0.861 & C & 16.7 & 0.02\textbf{*} & 4.1 & 178.50 \\ 
		16960 & 3 & 0.311 & 4.09 & 2.2 & 0.859 & 17.5 & 1.6 & Q & 14.3 & 0.227 & 4.1 & 5.79 \\ 
		20086 & 2.78 & 1.2 & 5.17 & 3.18 & 0.623 & 21.8 & 1.15 & C & 16.7 & 0.013\textbf{*} & 1.21 & - \\ 
		20790 & 3.09 & 1.207 & 4.28 & 2.74 & 0.56 & 8.3 & 0.872 & S & 16.4 & 0.1\textbf{*} & 1.56 & 2.79 \\ 
		20826 & 3.06 & 0.89 & 3.96 & 2.42 & 0.633 & 31.9 & 1.19 & Sq & 13.9 & 0.123\textbf{*} & 4.93 & 5.31 \\ 
		26760 & 3.04 & 1.26 & 4.44 & 2.85 & 0.558 & 11 & 0.939 & Xc & 15.3 & 0.042\textbf{*} & 5.39 & 5.30 \\ 
		52762 & 3.02 & 0.844 & 3.99 & 2.42 & 0.651 & 33.9 & 2.44 & D & 14.8 & 0.052\textbf{*} & 6.72 & 12.08 \\ 
		53319 & 2.99 & 0.976 & 4.47 & 2.73 & 0.642 & 13.8 & 0.855 & C & 15.2 & 0.03\textbf{*} & 7 & 136 \\ 
		53430 & 3.08 & 1.252 & 2.95 & 2.1 & 0.403 & 60.4 & 2.55 & Sr & 16.6 & 0.27\textbf{*} & 1.36 & 9.58 \\ 
		65996 & 2.96 & 1.163 & 4.71 & 2.94 & 0.604 & 9.6 & 0.971 & X & 18.5 & 0.047 & 0.59 & 3.40 \\ 
		85490 & 2.66 & 1.275 & 6.25 & 3.76 & 0.661 & 2.6 & 0.298 & D & 14.7 & 0.048 & 6.69 & 9.06 \\ 
		112221 & 2.93 & 1.259 & 3.28 & 2.27 & 0.445 & 58.7 & 1.73 & Sr & 15.5 & 0.266 & 2.36 & 12.87 \\ 
		137427 & 2.98 & 0.955 & 3.94 & 2.45 & 0.61 & 39.2 & 1.18 & S & 15.2 & 0.211 & 2.64 & - \\ 
		139359 & 2.68 & 0.343 & 4.92 & 2.63 & 0.87 & 5.9 & 0.629 & D & 16.4 & 0.04\textbf{*} & 3.49 & - \\ 
		152664 & 3.03 & 0.721 & 4.33 & 2.53 & 0.715 & 3.4 & 1.01 & S & 19.7 & 0.211 & 0.34 & 17.38 \\ 
		154453 & 2.74 & 0.426 & 4.74 & 2.58 & 0.835 & 20.8 & 0.64 & Sr & 15.2 & 0.266 & 2.71 & 5.23 \\ 
		162011 & 3.02 & 1.139 & 4.52 & 2.83 & 0.598 & 4.6 & 0.884 & Sq & 16.5 & 0.243 & 1.49 & 3.00 \\ 
		162186 & 2.94 & 1.059 & 4.37 & 2.71 & 0.61 & 27.6 & 2.28 & Sq & 15 & 0.243 & 2.7 & - \\ 
		162740 & 3.05 & 1.222 & 3.7 & 2.46 & 0.503 & 39.2 & 1.75 & Sq & 18.4 & 0.243 & 0.56 & - \\ 
		163250 & 2.94 & 1.24 & 4.14 & 2.69 & 0.539 & 35 & 0.889 & Sv & 15.5 & 0.309 & 1.92 & - \\ 
		163732 & 2.65 & 0.835 & 4.66 & 2.75 & 0.696 & 44.6 & 0.869 & C & 15.4 & 0.05 & 2.47 & 151.10 \\ 
		169675 & 3.1 & 1.228 & 4.21 & 2.72 & 0.549 & 12.6 & 1.3 & S & 16.5 & 0.211 & 1.44 & - \\ 
		170502 & 2.75 & 0.297 & 4.66 & 2.48 & 0.88 & 10.4 & 0.494 & C & 17.2 & 0.05 & 1.08 & 7.20 \\ 
		172678 & 3.03 & 0.806 & 4.38 & 2.59 & 0.689 & 2.7 & 0.582 & Q & 18.7 & 0.227 & 0.51 & - \\ 
		188452 & 3.09 & 1.088 & 4.01 & 2.55 & 0.573 & 24.7 & 1.46 & Sv & 17.4 & 0.309 & 0.98 & 5.36 \\ 
		189552 & 3.07 & 1.19 & 3.93 & 2.56 & 0.535 & 30.1 & 2.13 & Sr & 16.8 & 0.266 & 1.12 & - \\ 
		206910 & 3 & 0.734 & 4.42 & 2.58 & 0.715 & 5 & 0.982 & Q & 17.1 & 0.227 & 1.06 & - \\ 
		214088 & 2.86 & 0.897 & 4.89 & 2.89 & 0.69 & 13.3 & 1.21 & Sq & 15.3 & 0.25\textbf{*} & 2.59 & 6.34 \\ 
		214869 & 2.95 & 0.984 & 4.71 & 2.85 & 0.654 & 2 & 0.617 & Sq & 16.4 & 0.29\textbf{*} & 1.38 & 102.24 \\ 
		215188 & 2.94 & 0.904 & 4.47 & 2.69 & 0.663 & 22.3 & 1.99 & V & 15.5 & 0.362 & 2.47 & 9.24 \\ 
		217796 & 2.82 & 1.046 & 4.44 & 2.74 & 0.619 & 37.5 & 0.991 & Sr & 17.1 & 0.266 & 0.98 & - \\ 
		229007 & 3.05 & 0.336 & 4.06 & 2.2 & 0.847 & 8.6 & 1.17 & Sr & 16.9 & 0.266 & 1.07 & - \\ 
		230979 & 2.97 & 1.115 & 4.62 & 2.87 & 0.611 & 11.2 & 0.682 & K & 16.5 & 0.207\textbf{*} & 1.47 & - \\ 
		242708 & 3.08 & 0.762 & 4.24 & 2.5 & 0.695 & 0.8 & 0.748 & D & 18.1 & 0.048 & 0.71 & 63.80 \\ 
		248083 & 2.73 & 1.274 & 5.37 & 3.32 & 0.616 & 23.3 & 1.78 & Xk & 16 & 0.126\textbf{*} & 1.88 & 4.66 \\ 
		248590 & 2.46 & 0.88 & 4.95 & 2.91 & 0.698 & 52.3 & 0.332 & X & 16.5 & 0.018\textbf{*} & 4.96 & - \\ 
		250706 & 2.83 & 0.888 & 5.03 & 2.96 & 0.7 & 6.9 & 0.61 & L & 18.5 & 0.149 & 0.69 & - \\ 
		267494 & 2.54 & 0.585 & 4.85 & 2.72 & 0.785 & 46.8 & 3.29 & X & 16 & 0.047 & 1.62 & 2.43 \\ 
		276049 & 3.08 & 0.981 & 3.48 & 2.23 & 0.561 & 47.3 & 2.59 & Ch & 16.4 & 0.07\textbf{*} & 3.5 & 3.29 \\ 
		285944 & 3.07 & 1.081 & 3.3 & 2.19 & 0.507 & 53.1 & 2.15 & V & 16.5 & 0.407\textbf{*} & 1.49 & 2.25 \\ 
		293054 & 2.95 & 0.589 & 4.47 & 2.53 & 0.767 & 6.1 & 0.492 & S & 18.4 & 0.211 & 0.65 & 5.31 \\ 
		297274 & 2.97 & 0.502 & 4.37 & 2.44 & 0.794 & 2 & 0.767 & Sq & 16.8 & 0.243 & 1.3 & 4.65 \\ 
		307005 & 2.57 & 0.716 & 5.07 & 2.89 & 0.752 & 39.2 & 2.26 & C & 18 & 0.03\textbf{*} & 1.93 & - \\ 
		322652 & 3.04 & 1.127 & 4.17 & 2.65 & 0.575 & 24.4 & 1.19 & S & 16.9 & 0.211 & 1.13 & 2.39 \\ 
		326732 & 3.1 & 1.163 & 4.26 & 2.71 & 0.571 & 6.6 & 1.1 & D & 17.6 & 0.048 & 0.9 & 9.44 \\ 
		329614 & 2.99 & 0.866 & 4.51 & 2.69 & 0.678 & 5.4 & 0.99 & Q & 17.6 & 0.227 & 0.9 & 3.28 \\ 
		331471 & 2.69 & 0.264 & 4.73 & 2.5 & 0.894 & 14.3 & 0.624 & Sv & 15.6 & 0.309 & 2.25 & 45.50 \\ 
		333888 & 3.01 & 1.151 & 4.51 & 2.83 & 0.593 & 9.2 & 0.827 & D & 16.7 & 0.048 & 1.36 & 5.32 \\ 
		345853 & 3.04 & 1.269 & 4.38 & 2.83 & 0.551 & 13.8 & 1.06 & Sr & 16.2 & 0.266 & 1.57 & 56.80 \\ 
		353938 & 3.08 & 1.219 & 4.31 & 2.77 & 0.559 & 9.5 & 1.38 & Sq & 18.1 & 0.243 & 0.61 & 2.45 \\ 
		355256 & 2.78 & 1.233 & 5.45 & 3.34 & 0.631 & 12.5 & 0.552 & T & 16.8 & 0.042 & 1.14 & 7.15 \\ 
		361071 & 3.04 & 1.087 & 4.16 & 2.62 & 0.586 & 24.4 & 1.91 & Sr & 15.4 & 0.266 & 2.47 & 4.09 \\ 
		362310 & 2.96 & 0.52 & 4.36 & 2.44 & 0.787 & 12.7 & 1.22 & S & 18 & 0.211 & 0.73 & - \\ 
		380188 & 3.1 & 1.136 & 4.24 & 2.69 & 0.578 & 9.9 & 1.15 & Xc & 18.6 & 0.037\textbf{*} & 1.32 & - \\ 
		394130 & 2.31 & 0.082 & 5.1 & 2.59 & 0.968 & 33.5 & 0.821 & R & 17.1 & 0.157\textbf{*} & 1.28 & - \\ 
		399433 & 2.86 & 0.593 & 4.72 & 2.66 & 0.777 & 4.4 & 0.747 & Xc & 18.6 & 0.036\textbf{*} & 1.33 & - \\ 
		401857 & 2.56 & 0.408 & 5.24 & 2.82 & 0.856 & 22 & 0.926 & D & 16 & 0.046\textbf{*} & 3.82 & - \\ 
		409204 & 2.91 & 1.212 & 4.6 & 2.91 & 0.583 & 24 & 1.31 & C & 18.7 & 0.05 & 0.54 & 5.12 \\ 
		410650 & 2.98 & 1.23 & 4.68 & 2.95 & 0.584 & 6.7 & 0.76 & C & 18.5 & 0.05 & 0.59 & 47.30 \\ 
		410778 & 2.41 & 0.815 & 5.01 & 2.91 & 0.72 & 54.5 & 2.57 & B & 18 & 0.048\textbf{*} & 0.75 & 8.00 \\ 
		413038 & 3.04 & 1.121 & 4.18 & 2.65 & 0.577 & 24 & 2.28 & Sq & 16.8 & 0.3\textbf{*} & 1.24 & 6.57 \\ 
		413192 & 2.78 & 1.112 & 5.38 & 3.25 & 0.657 & 9.1 & 0.543 & Xc & 16.8 & 0.043\textbf{*} & 2.8 & - \\ 
		414287 & 2.62 & 0.774 & 5.64 & 3.21 & 0.759 & 13.5 & 0.818 & Xc & 17.7 & 0.05 & 0.86 & 42.50 \\ 
		414586 & 2.84 & 1.166 & 5.18 & 3.18 & 0.633 & 8.3 & 0.498 & S & 16 & 0.709\textbf{*} & 1.88 & - \\ 
		416071 & 3 & 1.024 & 4.31 & 2.67 & 0.616 & 22.3 & 1.97 & Q & 17.9 & 0.227 & 0.78 & 17.70 \\ 
		417264 & 3.01 & 1.088 & 4.33 & 2.71 & 0.599 & 21.2 & 1.02 & Cg & 17.2 & 0.062\textbf{*} & 1.98 & - \\ 
		428086 & 2.99 & 1.063 & 4.49 & 2.78 & 0.617 & 14.2 & 1.04 & S & 16.3 & 0.211 & 1.63 & 20.00 \\ 
		429584 & 3.07 & 0.75 & 4.25 & 2.5 & 0.7 & 2.6 & 1.1 & S & 19.8 & 0.211 & 0.33 & 43.50 \\ 
		430439 & 2.94 & 1.206 & 4.71 & 2.96 & 0.592 & 15.6 & 0.573 & T & 19.7 & 0.042 & 0.33 & 14.92 \\ 
		433992 & 2.59 & 1.171 & 4.66 & 2.92 & 0.599 & 49.3 & 1.43 & D & 18 & 0.144\textbf{*} & 0.75 & 9.36 \\ 
		442037 & 2.82 & 0.916 & 4.94 & 2.93 & 0.687 & 17.6 & 0.842 & Cg & 19.2 & 0.063\textbf{*} & 0.7 & 2.07 \\ 
		448972 & 3.06 & 1.292 & 4.39 & 2.84 & 0.545 & 6 & 0.59 & C & 17.2 & 0.05 & 1.08 & 2.85 \\ 
		450160 & 2.71 & 0.808 & 4.06 & 2.43 & 0.668 & 57.3 & 2.93 & Sr & 16.7 & 0.266 & 1.36 & 14.51 \\ 
		451124 & 2.73 & 0.963 & 5.45 & 3.21 & 0.7 & 10 & 0.156 & Cg & 18 & 0.023\textbf{*} & 2.2 & - \\ 
		452561 & 2.8 & 1.111 & 5.33 & 3.22 & 0.655 & 8.2 & 0.406 & C & 17.5 & 0.04\textbf{*} & 0.99 & 3.34 \\ 
		455192 & 3.02 & 1.223 & 4.54 & 2.88 & 0.576 & 2.7 & 0.876 & B & 18.1 & 0.12 & 0.68 & 30.33 \\ 
		455426 & 2.59 & 0.197 & 4.87 & 2.53 & 0.922 & 6.8 & 0.481 & S & 18.6 & 0.211 & 0.55 & - \\ 
		461912 & 3.02 & 0.942 & 4.45 & 2.7 & 0.65 & 8.6 & 1 & C & 18.2 & 0.05 & 1.37 & - \\ 
		465402 & 2.4 & 0.103 & 5.07 & 2.59 & 0.96 & 10.5 & 0.879 & Sr & 17.4 & 0.266 & 0.87 & - \\ 
		466130 & 2.39 & 0.976 & 4 & 2.49 & 0.608 & 75.4 & 0.965 & R & 18.3 & 0.148 & 0.77 & - \\ 
		468583 & 2.98 & 0.857 & 4.53 & 2.69 & 0.682 & 6.4 & 0.959 & S & 17.9 & 0.211 & 0.76 & - \\ 
		471241 & 2.8 & 0.779 & 4.98 & 2.88 & 0.729 & 14.6 & 0.985 & D & 18.2 & 0.044\textbf{*} & 0.68 & 4.83 \\ 
		475665 & 2.96 & 1.129 & 4.71 & 2.92 & 0.614 & 4.7 & 0.286 & C & 17.1 & 0.05 & 1.13 & 17.51 \\ 
		481032 & 2.21 & 1.105 & 5.74 & 3.42 & 0.677 & 56.1 & 1.73 & C & 15.1 & 0.02\textbf{*} & 8.97 & - \\ 
		485652 & 2.71 & 0.809 & 5.22 & 3.01 & 0.732 & 20.1 & 1.07 & T & 16.5 & 0.042 & 1.49 & 4.61 \\ 
		488645 & 2.99 & 0.324 & 4.16 & 2.24 & 0.856 & 10.3 & 1.35 & Q & 18.4 & 0.227 & 0.58 & - \\ 
		488693 & 2.82 & 1.12 & 4.08 & 2.6 & 0.569 & 46.8 & 1.39 & D & 17 & 0.048 & 2.41 & - \\ 
		496001 & 2.98 & 1.027 & 4.35 & 2.69 & 0.618 & 23.5 & 1.63 & Sv & 17.5 & 0.309 & 0.76 & - \\ 
		500080 & 2.94 & 0.896 & 4.68 & 2.78 & 0.678 & 6 & 0.304 & Sr & 17.1 & 0.266 & 1.13 & 10.11 \\ 
		506437 & 2.73 & 0.887 & 5.39 & 3.14 & 0.717 & 10.3 & 0.936 & Xc & 18 & 0.129 & 0.93 & - \\ 
		523915 & 2.79 & 0.49 & 4.75 & 2.62 & 0.813 & 14.2 & 0.851 & K & 18.2 & 0.13 & 0.84 & - \\ 
		524516 & 2.82 & 1.171 & 4.84 & 3 & 0.61 & 26.2 & 1.27 & Cg & 17.1 & 0.063 & 2.01 & - \\ 
		1991 XB & 2.94 & 1.236 & 4.68 & 2.96 & 0.583 & 16.3 & 0.694 & K & 18.8 & 0.13 & 0.64 & - \\ 
		1998 GL10 & 2.79 & 1.053 & 5.29 & 3.17 & 0.668 & 8.7 & 0.712 & X & 19 & 0.047 & 0.47 & 5.93 \\ 
		1998 HT31 & 3.05 & 0.769 & 4.29 & 2.53 & 0.696 & 6.8 & 1.13 & C & 20.8 & 0.05 & 0.41 & - \\ 
		1999 DB2 & 2.9 & 1.179 & 4.89 & 3.04 & 0.612 & 11 & 0.975 & Sq & 19 & 0.243 & 0.43 & - \\ 
		1999 SE10 & 2.85 & 1.226 & 5.21 & 3.22 & 0.619 & 6.9 & 0.246 & D & 20 & 0.048 & 0.61 & - \\ 
		2000 CN33 & 3.1 & 1.075 & 4.13 & 2.6 & 0.587 & 17.7 & 0.999 & X & 19.4 & 0.047 & 0.81 & - \\ 
		2000 GV127 & 2.95 & 1.094 & 4.59 & 2.84 & 0.615 & 17.9 & 0.365 & S & 19.2 & 0.211 & 0.42 & - \\ 
		2000 WJ63 & 3.02 & 1.159 & 4.51 & 2.84 & 0.591 & 2.8 & 0.823 & Sq & 20.9 & 0.243 & 0.18 & - \\ 
		2001 EC & 2.91 & 0.609 & 4.59 & 2.6 & 0.766 & 0.6 & 0.748 & V & 18.5 & 0.362 & 0.44 & - \\ 
		2001 SJ262 & 2.98 & 1.252 & 4.64 & 2.95 & 0.575 & 10.8 & 0.888 & Ch & 19.9 & 0.177\textbf{*} & 0.33 & - \\ 
		2001 UC5 & 2.88 & 1.005 & 4.47 & 2.74 & 0.633 & 30.4 & 1.29 & D & 21.3 & 0.048 & 0.31 & 0.03 \\ 
		2001 UU92 & 2.8 & 1.058 & 5.29 & 3.17 & 0.667 & 5.4 & 0.472 & D & 20.1 & 0.048 & 0.58 & - \\ 
		2002 AQ2 & 3.07 & 1.046 & 4.28 & 2.66 & 0.607 & 11.7 & 0.849 & S & 18.6 & 0.211 & 0.55 & - \\ 
		2002 XO14 & 2.98 & 0.985 & 4.59 & 2.79 & 0.647 & 2.6 & 0.533 & Sr & 22.1 & 0.266 & 0.1 & - \\ 
		2003 RS1 & 2.99 & 0.994 & 4.56 & 2.78 & 0.642 & 2.7 & 0.864 & X & 21.7 & 0.047 & 0.28 & - \\ 
		2003 XM & 2.78 & 1.031 & 5.37 & 3.2 & 0.678 & 5.6 & 0.034 & T & 19.1 & 0.042 & 0.98 & - \\ 
		2006 UN216 & 2.84 & 1.236 & 5.2 & 3.22 & 0.616 & 11.8 & 0.15 & B & 18.9 & 0.12 & 0.64 & - \\ 
		2008 JY30 & 3.06 & 0.575 & 4.18 & 2.37 & 0.758 & 10.9 & 1.27 & Q & 18.9 & 0.227 & 0.46 & - \\ 
		2008 SV11 & 2.96 & 0.717 & 4.51 & 2.61 & 0.725 & 8.3 & 0.882 & X & 18.4 & 0.047 & 0.62 & 32.40 \\ 
		2009 QJ9 & 2.99 & 1.032 & 4.41 & 2.72 & 0.621 & 19.3 & 1.13 & S & 19.3 & 0.211 & 0.4 & - \\ 
		2009 SV17 & 3.09 & 1.118 & 4.29 & 2.7 & 0.586 & 3.4 & 0.995 & Sr & 19.6 & 0.266 & 0.34 & 17.34 \\ 
		2010 GT7 & 2.95 & 0.852 & 4.59 & 2.72 & 0.687 & 9.1 & 0.77 & B & 20.2 & 0.326\textbf{*} & 0.21 & - \\ 
		2011 BE38 & 2.97 & 0.757 & 4.5 & 2.63 & 0.712 & 8.2 & 0.866 & D & 18.3 & 0.048 & 1.33 & - \\ 
		2011 LJ19 & 3.1 & 0.993 & 4.25 & 2.62 & 0.621 & 1.5 & 1.14 & S & 21.3 & 0.211 & 0.16 & - \\ 
		2011 YB40 & 3 & 1.041 & 4.07 & 2.56 & 0.593 & 32.7 & 1.86 & Q & 19.2 & 0.227 & 0.4 & - \\ 
		2012 LZ1 & 3.06 & 1.047 & 4.07 & 2.56 & 0.59 & 26.1 & 1.15 & X & 19.9 & 0.047 & 0.8 & 12.87 \\ 
		2012 RV16 & 2.9 & 0.91 & 4.63 & 2.77 & 0.671 & 20.2 & 1.76 & Sr & 19.6 & 0.266 & 0.31 & - \\ 
		2012 XJ134 & 2.98 & 0.673 & 4.46 & 2.57 & 0.738 & 2.5 & 1.04 & X & 21.5 & 0.047 & 0.31 & - \\ 
		2013 EO126 & 3.09 & 0.5 & 4.09 & 2.3 & 0.782 & 3 & 1.2 & S & 19.9 & 0.211 & 0.3 & - \\ 
		2013 SW24 & 3.03 & 0.932 & 4.42 & 2.67 & 0.652 & 6.1 & 0.935 & Sr & 21 & 0.266 & 0.19 & 7.09 \\ 
		2014 RL12 & 2.88 & 0.903 & 4.62 & 2.76 & 0.673 & 23.8 & 1.91 & Sv & 17.9 & 0.256\textbf{*} & 0.78 & 5.21 \\ 
		2014 UF206 & 2.93 & 1.108 & 3.74 & 2.43 & 0.543 & 48.1 & 1.45 & Cb & 18.8 & 0.028\textbf{*} & 0.52 & 6.83 \\ 
		2015 CA1 & 2.96 & 1.097 & 4.64 & 2.87 & 0.617 & 11.2 & 0.422 & T & 20.6 & 0.042 & 0.23 & 2.95 \\ 
		2015 OL35 & 3.09 & 0.921 & 4.25 & 2.59 & 0.644 & 4.5 & 1.11 & R & 18.1 & 0.148 & 0.75 & 5.35 \\ 
		2015 XB379 & 2.82 & 1.155 & 5.24 & 3.2 & 0.639 & 8.4 & 0.001 & S & 19.1 & 0.211 & 0.45 & 48.20 \\ 
		2016 LX48 & 2.75 & 0.994 & 5.45 & 3.22 & 0.692 & 5.6 & 0.054 & Cg & 19.3 & 0.063 & 0.41 & 5.68 \\ 
		2016 OJ1 & 3.03 & 1.092 & 4.49 & 2.79 & 0.608 & 4.6 & 0.982 & Sr & 21.6 & 0.266 & 0.12 & - \\ 
		2016 PB8 & 3.07 & 1.092 & 4.34 & 2.72 & 0.598 & 7.3 & 1.18 & X & 22.5 & 0.047 & 0.19 & - \\ 
		2016 PR8 & 2.88 & 0.918 & 4.76 & 2.84 & 0.677 & 16.3 & 1.09 & Cb & 18.8 & 0.043 & 0.52 & 16.17 \\ 
		2017 MB1 & 3.08 & 0.589 & 4.16 & 2.37 & 0.752 & 8.5 & 1.42 & V & 18.8 & 0.362 & 0.52 & 6.69 \\ 
		2017 YE5 & 2.88 & 0.819 & 4.82 & 2.82 & 0.709 & 6.2 & 0.422 & X & 19.2 & 0.047 & 0.43 & 20.60 \\ 
		2018 MM8 & 3.06 & 1.154 & 4.2 & 2.67 & 0.569 & 19.5 & 1.36 & Sr & 19.2 & 0.266 & 0.43 & 5.05 \\ 
		2018 TT1 & 2.85 & 1.091 & 4.41 & 2.75 & 0.603 & 35.8 & 1.39 & Sv & 18.5 & 0.309 & 0.48 & - \\ 
		2019 HC & 2.94 & 1.199 & 4.14 & 2.67 & 0.551 & 35.3 & 0.847 & A & 17.2 & 0.191 & 1.08 & 1.26 \\ 
		333P & 0.42 & 1.115 & 7.33 & 4.22 & 0.736 & 131.9 & 0.358 & D & 15 & 0.048 & 6.07 & 1.14 \\ 
		\hline
	\end{longtable}
	\twocolumn
	% Maybe some additional info will be required here  
	
	%%%%%%%%%%%%%%%%%%%%%%%%%%%%%%%%%%%%%%%%%%%%%%%%%%

	% Don't change these lines
	\bsp	% typesetting comment
	\label{lastpage}
\end{document}